\documentclass[11pt]{article}
\usepackage[T1]{fontenc}
\usepackage{lmodern}
\usepackage{microtype}
\usepackage{amsmath,amssymb,mathtools,bm,amsthm}
\usepackage{mathrsfs} % distinct calligraphic kernel in the root-of-unity proof
\usepackage{booktabs,longtable,array,tabularx}
\usepackage{graphicx}
\usepackage{authblk} % Article-class compatibility for journal-style affiliations
\let\affiliation\affil

\usepackage{tikz}
\usepackage{pgfplots}
\usepackage{float}
\pgfplotsset{compat=1.18}
\usepackage{cite} % JSTAT: numeric citations sorted and compressed
\usepackage[hidelinks]{hyperref}
\usepackage{xcolor}
\usepackage{enumitem}
\usepackage{verbatim}
\newcommand{\dd}{\,\mathrm d}
\newcommand{\EE}{\mathbb E}

\newcommand{\cZ}{\mathcal Z}
\newcommand{\one}{\mathbf 1}

\newtheorem{theorem}{Theorem}[section]

\newtheorem{lemma}[theorem]{Lemma}
\newtheorem{conjecture}[theorem]{Conjecture}
\theoremstyle{remark}

\numberwithin{equation}{section}
\title{High-temperature expansion of the inverse-square Ising chain in a magnetic field: exact coefficients and finite-size reconstruction}
\author{M. A. Rajabpour}
\affiliation{Instituto de F\'isica, Universidade Federal Fluminense, Av. Gal. Milton Tavares de Souza s/n, Gragoat\'a, 24210-346 Niter\'oi, RJ, Brazil}
\affiliation{Theoretical Physics III, Center for Electronic Correlations and Magnetism, Institute of Physics, University of Augsburg, D-86135 Augsburg, Germany}
\date{}
\begin{document}
\maketitle
\vspace{-3\baselineskip}

\begin{abstract}
The one-dimensional Ising chain with inverse-square interactions is a paradigmatic long-range system: it exhibits the unusual Thouless transition, with a discontinuity in the spontaneous magnetization, yet its full field-dependent thermodynamics remains unknown in closed form. We develop the high-temperature expansion of its free energy while retaining the dependence on a uniform magnetic field at every order. We prove that, after a natural rescaling of the coupling, each coefficient is a rational polynomial in the squared hyperbolic tangent of the reduced field. Remarkably, at each order $n$ the field-dependent free-energy coefficient of a finite, periodically repeated ring is an exact polynomial of degree at most $n$ in $L^{-2}$. The infinite-chain coefficient is its constant term and can therefore be reconstructed from $n+1$ finite rings without numerical extrapolation. Using this reconstruction, we determine the full field-dependent expansion through order 26: 26 polynomials containing 377 rational coefficient entries. We further derive exact formulas, valid at every order, for the coefficients of the two highest powers of the field variable using matching determinants and a cycle-Laplacian reduction. These results reveal an unexpected algebraic structure in the microscopic thermodynamics of the inverse-square chain and provide high-order information about its magnetic response.
\end{abstract}


% A concise contents page helps readers navigate the long proofs and appendices.
% Only main section and appendix section titles are listed.
\clearpage
\setcounter{tocdepth}{1}
\begingroup
\small
\tableofcontents
\endgroup
\clearpage

\section{Introduction}
The one-dimensional Ising chain with ferromagnetic inverse-square interactions occupies a special place in statistical mechanics. Short-range Ising chains do not order at finite temperature, whereas sufficiently slowly decaying ferromagnetic interactions can sustain an ordered phase even in one dimension \cite{Ruelle1968,Dyson1969}. The interaction proportional to $1/r^2$ lies at the marginal boundary between these regimes. A discontinuous onset of spontaneous magnetization at this borderline decay was first predicted in the original work on the Thouless effect \cite{Thouless1969}. The existence of a finite-temperature ordered phase was subsequently established rigorously \cite{FrohlichSpencer1982}, as was the discontinuity of the spontaneous magnetization \cite{AizenmanEtAl1988}. The resulting Thouless transition is accompanied by unusual correlations and critical behavior \cite{ImbrieNewman1988,LuijtenMessingfeld2001}. Despite this substantial progress, the microscopic inverse-square chain has no complete closed-form free energy or field-dependent equation of state. It remains one of the paradigmatic analytically unsolved models of one-dimensional classical statistical mechanics.

The physical origin of this behavior connects the model to several broader themes. Reversing a segment of spins produces two domain walls whose interaction energy grows logarithmically with their separation. This leads to a Coulomb-gas description closely related to formulations of the Kondo problem \cite{AndersonYuval1969,YuvalAnderson1970,AndersonYuvalHamann1970,AndersonYuval1971}. Renormalization-group arguments further illuminate the marginal character of the inverse-square interaction and the unusual structure of the transition \cite{Kosterlitz1976,Cardy1981,BhattacharjeeEtAl1981}. The chain thus provides a particularly concrete setting in which long-range ordering, logarithmic defects, and unconventional criticality meet.

Over the years, the model has been investigated using numerical calculations, bounds, finite-range scaling, and a variety of lattice approximations \cite{NagleBonner1970,Ojo1973,Doman1981,Vigfusson1986,UzelacGlumac1988,GlumacUzelac1989,MonroeEtAl1990,Monroe1992,Mainieri1992,Cannas1995,Pires1996,FukuiTodo2009,MartinezHerrera2022}. Its transition temperature continues to be examined with modern analytical and numerical methods \cite{TowerEtAl2026}. Rigorous studies have sharpened our understanding of ordering, correlations, and phase separation in one-dimensional long-range systems \cite{Imbrie1982,CassandroMerolaPiccoRozikov2014,DuminilCopinGarbanTassion2024,AffonsoEtAl2024,BissacotCorsini2026}. Parallel developments in the continuum and field-theoretic treatment of long-range criticality place the model within a broader family of related problems \cite{BenedettiGurauHarribeySuzuki2020,BenedettiGurauHarribey2025,BenedettiLauriaMazacVanVliet2025,BenedettiLauriaMazacVanVliet2026,PagniEtAl2026,GhoshEtAl2026}. These approaches have greatly clarified the physics of the transition, while a full solution of the field-dependent lattice thermodynamics has remained out of reach.

A complementary route is to seek exact information directly from the microscopic high-temperature expansion. Earlier work on the inverse-square chain obtained the zero-field susceptibility series through eighth order and the specific-heat series through ninth order \cite{Matvienko1985}; the wider role of high-temperature series in lattice statistical mechanics is reviewed in \cite{OitmaaHamerZheng2006}. The present goal is different from simply extending a single zero-field series. At each perturbative order, the full magnetic-field dependence of the free energy determines the equation of state and an entire hierarchy of nonlinear magnetic responses. The same coefficients can also be continued to imaginary magnetic field, providing a microscopic starting point for studying analyticity in the setting of Lee--Yang theory \cite{LeeYang1952,YangLee1952,GlumacUzelac1991}. For a model whose complete thermodynamics is unknown, it is therefore natural to ask how much of this field-dependent expansion can be obtained exactly.

We consider the infinite-chain Hamiltonian
\begin{equation}
 H=-J\sum_{i<j}\frac{\sigma_i\sigma_j}{(i-j)^2}
   -H_{\mathrm{ext}}\sum_i\sigma_i,
\end{equation}
where $\sigma_i=\pm1$, $J>0$, and $H_{\mathrm{ext}}$ is the magnetic-field energy per spin. We use $g=\beta J$, $x=\pi^2g$, and the reduced magnetic field $h=\beta H_{\mathrm{ext}}$. The dimensionless free energy per spin, $\phi=-\beta f$, has the high-temperature expansion
\begin{equation}
 \begin{aligned}
 \phi(x,h)&=\lim_{L\to\infty}\frac{1}{L}\log Z_L\\
 &=\log(2\cosh h)+\sum_{n\ge1}\mathcal P_n(y)x^n,
 \qquad y=\tanh^2h.
 \end{aligned}
\end{equation}
Here $Z_L$ is the finite-volume partition function with spin-independent self-image terms omitted, and the thermodynamic limit is taken coefficientwise in $x$. For finite rings we sum the inverse-square interaction over periodic images, rather than imposing a minimum-image cutoff. Our objects of interest are the complete field polynomials $\mathcal P_n(y)$.

The first main result is an exact statement valid at every order:
\begin{equation}
 \mathcal P_n(y)\in\mathbb Q[y],\qquad
 \deg\mathcal P_n\le n.
 \label{eq:intro-rational}
\end{equation}
Finite-degree dependence on $\tanh h$ is familiar from Ising graph expansions. The special property of the inverse-square chain is that every spatial coefficient becomes rational when the coupling is expressed in terms of $x=\pi^2\beta J$. Here $\mathbb Q$ denotes the rational numbers. The proof combines a connected-cumulant construction with independent local magnetic sources and the exact Fourier transform of the periodized interaction, which is expressed through a periodic Bernoulli polynomial. This reduces the spatial sums to integrals of piecewise polynomials over rational polyhedra. The underlying source-cumulant, cosecant-sum, Bernoulli, and determinant techniques have a substantial history \cite{FarrellMoritaMeijer1966,LeonovShiryaev1959,GrabnerProdinger2007,CvijovicSrivastava2007,CvijovicSrivastava2012,Dowker2015,Szenes1998,SzenesVergne2003,BoysalVergne2012,IshikawaEtAl2005}. Related inverse-square kernels have also appeared in quantum-chain measurement statistics \cite{Rajabpour2026Renyi,Rajabpour2026Shannon}; here it is their role in the thermal, field-dependent expansion that yields the rationality theorem.

The field polynomials also satisfy the exact saturation condition $\mathcal P_n(1)=0$ for every $n\ge2$. In the fully polarized limit, spin fluctuations disappear and the interaction contributes to the free energy only linearly in the coupling.

The second main result is perhaps more surprising: the dependence on system size itself has an exact, finite algebraic form. We prove that, for a periodized ring of $L$ spins, the coefficient at every high-temperature order $n$ can be written as
\begin{equation}
 \mathcal P_{n,L}(y)=\sum_{j=0}^{n}\frac{c_{n,j}(y)}{L^{2j}},
 \qquad c_{n,j}(y)\in\mathbb Q[y].
 \label{eq:intro-ring}
\end{equation}
The sum terminates exactly: there are no additional finite-size corrections at that order. Since the infinite-chain coefficient is $\mathcal P_n(y)=c_{n,0}(y)$, it can be determined from any $n+1$ distinct ring lengths, without numerical extrapolation. This holds at arbitrary perturbative order and therefore provides, in principle, a systematic route to increasingly high-order exact thermodynamic coefficients. The proof combines the Bernoulli Fourier representation, weighted integral-grid reciprocity, and a complementary position-space root-of-unity summation, with a sharp bound for connected graph amplitudes. The complete mathematical argument is given in Appendix~\ref{app:A-reconstruction}. In practice, the cost of the finite-ring calculations grows rapidly with $n$. We use this exact reconstruction to determine the full field-dependent expansion through order 26, comprising 26 polynomials with $\sum_{n=1}^{26}(n+1)=377$ rational coefficient entries, rather than a single coefficient at each order.

An important consequence is that the same finite-ring calculation determines the zero-field susceptibility directly. In terms of the field polynomials,
\begin{equation}
 \chi(x)=\left.\frac{\partial^2\phi(x,h)}{\partial h^2}\right|_{h=0}
 =1+\sum_{n\ge1}a_n^{(\chi)}x^n,
 \qquad a_n^{(\chi)}=2\mathcal P_n'(0),
 \label{eq:intro-susceptibility}
\end{equation}
where the prime denotes differentiation with respect to $y$. The sharp finite-size theorem applies to these coefficients as well: $n+1$ distinct ring lengths determine each thermodynamic susceptibility coefficient exactly. Hence the susceptibility expansion can be extended to higher orders without reconstructing the complete field polynomial at each order.

A further result is that two entire sequences within this field-coefficient array have simple closed forms. Writing $\mathcal P_n(y)=\sum_{m=0}^{n}p_{n,m}y^m$, we prove
\begin{equation}
 p_{n,n}=\frac{(-1)^{n+1}}{2n(2n+1)},\qquad n\ge1,
 \label{eq:intro-top}
\end{equation}
\begin{equation}
 p_{n,n-1}=(-1)^n\frac{5n-4}{9(2n-1)(2n+1)},\qquad n\ge2.
 \label{eq:intro-next}
\end{equation}
The first follows from a weighted-matching determinant, and the second from deleted matching minors and a cycle-Laplacian reduction. Both formulas hold at arbitrary perturbative order. The computed coefficients suggest additional regularities, including alternating signs across the nonconstant powers of $y$ and rational formulas along further subleading diagonals. These patterns motivate the conjectures discussed below.

The combination of exact theorems, all-order coefficient formulas, and the high-order field-polynomial array provides a new microscopic description of the inverse-square chain. Besides the free energy itself, the expansion gives access to nonlinear magnetic responses and to exploratory analytic continuation in the magnetic field. Section~2 introduces the exact finite-ring Fourier kernel and the connected-source expansion. Section~3 proves rationality and the sharp finite-size reconstruction theorem. Section~4 presents the coefficient array, and Section~5 derives the two all-order coefficient formulas. Section~6 discusses magnetic response and imaginary-field continuation; the appendices contain the detailed arguments and computational checks.

\section{Fourier interaction kernel and connected magnetic sources}\label{sec:tools}
The calculation of the field-dependent high-temperature expansion rests on two complementary ingredients. The first is an exact finite-ring representation of the inverse-square interaction, obtained by summing over periodic images. Its discrete Fourier transform takes a particularly simple form in terms of Bernoulli polynomials, allowing the spatial sums arising in the expansion to be treated analytically. The second is a connected expansion of the partition function in the presence of independent local magnetic sources. This construction provides a systematic way to obtain the free-energy coefficients while retaining their full magnetic-field dependence and correctly accounting for repeated bonds and shared lattice sites. In this section, we develop these two ingredients and illustrate how they combine in the calculation of the first high-temperature coefficient.

\subsection{Periodized inverse-square interaction and its Fourier transform}\label{sec:model}
We periodize the inverse-square interaction by summing over all translated images of an $L$-site cell. This differs from the minimum-image convention and from a sharp distance cutoff. For distinct sites separated by $r$, the standard partial-fraction identity gives
\begin{equation}
 \sum_{m\in\mathbb Z}\frac{1}{(r+mL)^2}
 =\frac{\pi^2}{L^2}\csc^2\!\left(\frac{\pi r}{L}\right),
 \qquad 1\le r\le L-1.
\end{equation}
With $x=\pi^2\beta J$, the dimensionless interaction kernel is therefore
\begin{equation}
 A_L(r)=\frac{1}{L^2}\csc^2\!\left(\frac{\pi r}{L}\right),
 \qquad 1\le r\le L-1.
\end{equation}
The images of a spin with itself contribute only a spin-independent constant. Retaining these diagonal image terms would shift the finite-ring logarithm of the partition function per spin by $x/(6L^2)$. We omit them throughout, so $A_L(0)=A_{ii}=0$. This convention fixes the finite-size coefficients without affecting their thermodynamic limits.

For integer momentum $k=0,\ldots,L-1$, the discrete Fourier transform is
\begin{equation*}
 \widehat A_L(k)=\sum_{r=0}^{L-1}A_L(r)e^{-2\pi i kr/L}.
\end{equation*}
Its exact value is
\begin{equation}
 \widehat A_L(k)=2B_2(k/L)-\frac{1}{3L^2},
 \qquad B_2(u)=u^2-u+\frac16.
 \label{eq:A-symbol-main}
\end{equation}
For completeness, this follows from a short finite-difference calculation. Set $S_L(k)=L^2\widehat A_L(k)$. The identity
$e^{2\pi ir/L}+e^{-2\pi ir/L}-2=-4\sin^2(\pi r/L)$
implies
\begin{equation}
 S_L(k+1)-2S_L(k)+S_L(k-1)=4,
 \qquad 1\le k\le L-1.
\end{equation}
The initial value $S_L(0)=(L^2-1)/3$ follows from the finite cosecant sum
$\sum_{r=1}^{L-1}\csc^2(\pi r/L)=(L^2-1)/3$ \cite{GrabnerProdinger2007,CvijovicSrivastava2007}. One way to obtain this standard identity is to subtract the $r=0$ term from
$\sum_{r=0}^{L-1}\csc^2(v+\pi r/L)=L^2\csc^2(Lv)$ and let $v\to0$. Together with $L$-periodicity, the difference equation yields
\begin{equation}
 S_L(k)=\frac{L^2-1}{3}-2k(L-k),
 \qquad 0\le k\le L-1,
\end{equation}
which gives the stated Fourier symbol after division by $L^2$. No large-$L$ approximation is involved.

Two consequences will be used below. At zero momentum, $\widehat A_L(0)=\sum_{r\ne0}A_L(r)=(1-L^{-2})/3$ fixes the first expansion coefficient. For general momenta, the product of Fourier symbols along the edges of an interaction graph gives the discrete sum in Eq.~\eqref{eq:site-fourier-main}; the periodic Bernoulli factors become the rational graph integrals in Eq.~\eqref{eq:torus-graph-main}. The explicit $L^{-2}$ correction enters the finite-size analysis. Together with position-space root-of-unity summation, the Bernoulli representation establishes the exact even-power structure in Section~3 and Appendix~\ref{app:A-reconstruction}.

\subsection{Recursion for the connected partition function with site-dependent fields}
The Fourier symbol evaluates the interactions, but it does not by itself distinguish connected from disconnected contributions.  This distinction is essential when calculating $\log Z$: disconnected products must cancel order by order.  Keeping an independent magnetic source at each site makes the cancellations local and gives an exact recursion that can be used without enumerating connected graphs individually \cite{FarrellMoritaMeijer1966}.

Let $h_i$ be independent dimensionless sources, $A_{ij}=A_L(i-j)$ for $i\ne j$, and $A_{ii}=0$.  We normalize the finite-ring partition function by the zero-field spin count and write
\begin{equation}
 \mathcal Z_L(x,\bm h)=2^{-L}\sum_{\{\sigma\}}\exp\!\left[
 \frac{x}{2}\sum_{i,j=0}^{L-1}A_{ij}\sigma_i\sigma_j
 +\sum_{i=0}^{L-1}h_i\sigma_i\right],\qquad
 W_L=\log\mathcal Z_L.
 \label{eq:finite-source-main}
\end{equation}
At $x=0$, $W_L(0,\bm h)=\sum_i\log\cosh h_i$.  At equal sources, the logarithm of the physical finite-ring partition function per spin is
\begin{equation}
 \begin{aligned}
 \phi_L(x,h)&=\log2+\frac1L W_L(x,h\bm1)\\
 &=\log(2\cosh h)+\sum_{n\ge1}\mathcal P_{n,L}(y)x^n,
 \qquad y=\tanh^2h.
 \end{aligned}
 \label{eq:finite-coeff-definition}
\end{equation}
This defines the finite-ring field polynomial $\mathcal P_{n,L}$; the bulk coefficient $\mathcal P_n$ is its coefficientwise limit as $L\to\infty$. In the next equations $\partial_i$ denotes $\partial/\partial h_i$.  Differentiating the exponential with respect to $x$ inserts a pair of spins, which is equivalent to two source derivatives.  Thus the partition function obeys the exact heat equation
\begin{equation}
 \partial_x\mathcal Z_L=\frac12\sum_{ij}A_{ij}\partial_i\partial_j\mathcal Z_L,
\end{equation}
so
\begin{equation}
 \partial_xW_L=\frac12\sum_{ij}A_{ij}\left(\partial_i\partial_jW_L+\partial_iW_L\partial_jW_L\right).
 \label{eq:W-main}
\end{equation}
Expanding $W_L=\sum_{n\ge0}x^nW_{n,L}$ makes the connectedness bookkeeping explicit:
\begin{equation}
 (n+1)W_{n+1,L}=\frac12\sum_{ij}A_{ij}
 \left[\partial_i\partial_jW_{n,L}+
 \sum_{r=0}^{n}(\partial_iW_{r,L})(\partial_jW_{n-r,L})\right].
 \label{eq:W-rec-main}
\end{equation}
The initial condition supplies the local cumulants $\kappa_m(t)=\partial_h^m\log\cosh h$ at $t=\tanh h$.  They obey $\kappa_{m+1}(t)=(1-t^2)\partial_t\kappa_m(t)$, starting from $\kappa_1=t$, so they are integer polynomials in $t$ with a definite parity.  Applying the recursion before imposing $h_i=h$ preserves repeated-bond and shared-site terms. In the connected expansion, a physical lattice site can support several local cumulant blocks; treating its full bond degree as a single cumulant would in general miss contributions. Once the spatial sums have been evaluated, the local cumulant polynomials give the field dependence. Appendix~\ref{app:A-source} derives the second-order coefficient explicitly and makes this distinction concrete.

To see both ingredients at work, the first nonzero term of Eq.~\eqref{eq:W-rec-main} is
\begin{equation}
 W_{1,L}(\bm h)=\frac12\sum_{i,j}A_{ij}\tanh h_i\tanh h_j,
\end{equation}
because the off-diagonal mixed derivatives of $W_{0,L}=\sum_i\log\cosh h_i$ vanish. Setting all $h_i=h$ gives, with $y=\tanh^2h$,
\begin{equation}
 \mathcal P_{1,L}(y)=\frac{y}{2L}\sum_{i,j}A_{ij}
  =\frac{y}{2}\widehat A_L(0)=\frac{y}{6}(1-L^{-2}).
 \label{eq:source-first-order}
\end{equation}
Thus the sources supply the field factor and the Fourier spectrum supplies its spatial coefficient. At second order both repeated bonds and distinct bonds meeting at a site contribute; Appendix~\ref{app:A-source} evaluates them separately, and Eq.~\eqref{eq:P2-ring-example} records the resulting finite-ring polynomial. For general $n$, connected \emph{cumulant-block graphs} provide the bridge to the spatial amplitudes discussed in Section~3 and Appendix~\ref{app:A-reconstruction}; their block vertices must not automatically be identified with distinct physical lattice sites.

\section[Rational field coefficients and finite-size reconstruction]{Rational high-temperature coefficients and exact finite-size reconstruction}
The kernel in Eq.~\eqref{eq:A-symbol-main} and source recursion in Eq.~\eqref{eq:W-rec-main} now enter the same graph expansion: the sources determine which connected graphs occur and their local field factors, while Fourier momentum conservation evaluates the spatial weights. We use this to answer two questions. Are the coefficients rational in the normalization $x=\pi^2\beta J$? Can the infinite-volume coefficients be obtained from finitely many rings? The first answer uses integrals of piecewise-polynomial Bernoulli functions; the second uses polynomiality of their finite-grid sums.

\subsection{Uniform-field theorem}
Throughout, $[v^m]F(v)$ denotes the coefficient of $v^m$ in the expansion of $F(v)$.
The coefficients $\mathcal P_{n,L}(y)$ were defined from the uniform-source partition function in Eq.~\eqref{eq:finite-coeff-definition}, and their coefficientwise limits are $\mathcal P_n(y)$. The field-degree bound comes from the $2n$ bond endpoints at order $n$; rationality additionally requires the special inverse-square spatial kernel.
\begin{theorem}[Rational uniform-field coefficients]\label{thm:field}
At every fixed perturbative order $n\ge1$, the coefficient $\mathcal P_n$ in the expansion of $\phi=-\beta f$ satisfies
\begin{equation}
 \mathcal P_n(y)\in\mathbb Q[y],\qquad \deg\mathcal P_n\le n.
 \label{eq:field-closure-main}
\end{equation}
Moreover $\mathcal P_1(y)=y/6$ and $\mathcal P_n(1)=0$ for $n\ge2$.
\end{theorem}
\begin{proof}
\emph{Connected expansion.} Let $T=\frac12\sum_{i,j}A_{ij}\sigma_i\sigma_j$ and
$\langle\cdot\rangle_{0,\bm h}$ denote the independent-spin measure at
$x=0$. Since
$W_L(x,\bm h)-W_L(0,\bm h)=\log\langle e^{xT}\rangle_{0,\bm h}$,
its order-$n$ coefficient is $\operatorname{cum}_0(T,\ldots,T)/n!$.
Label the $2n$ endpoints of the $n$ bond occurrences, and let
$\eta=\{\{1,2\},\ldots,\{2n-1,2n\}\}$ be their bond partition.
For a partition $\pi$ of these endpoints, write $G_\pi$ for the graph
whose vertices are the blocks of $\pi$ and whose edges are the pairs in
$\eta$. The connected-cluster organization is standard in Ising expansions \cite{FarrellMoritaMeijer1966}. The product-cumulant identity \cite{LeonovShiryaev1959} retains precisely the endpoint partitions that connect all $n$ bond occurrences, expressed compactly by $\pi\vee\eta=\mathbf1$, where $\pi\vee\eta$ is the partition generated jointly by $\pi$ and $\eta$, and $\mathbf1$ denotes the single-block partition; it gives the exact finite-ring formula
\begin{equation}
 \mathcal P_{n,L}(t^2)=\frac{1}{2^n n!}
 \sum_{\substack{\pi\in\Pi(2n)\\\pi\vee\eta=\mathbf1}}
 \left[\prod_{B\in\pi}\kappa_{|B|}(t)\right]
 \mathcal A_{G_\pi,L},\qquad t=\tanh h,
 \label{eq:connected-partition-main}
\end{equation}
where $\Pi(2n)$ denotes the set of endpoint partitions and
\begin{equation}
 \mathcal A_{G,L}=\frac1L
 \sum_{\bm s\in(\mathbb Z/L\mathbb Z)^{V(G)}}
 \prod_{e=(u,v)\in E(G)}A_L(s_u-s_v).
 \label{eq:block-spatial-main}
\end{equation}
The assignments of site labels $s_v$ are \emph{unrestricted}: distinct
cumulant blocks may occupy the same physical site. A partition producing
a self-loop contributes zero because $A_L(0)=0$. In particular, the
normalization $2^{-n}n!^{-1}$ and the treatment of repeated bonds are
fixed by the original partition function, not by graph symmetry
conventions. Appendix~\ref{app:A-reconstruction} derives this formula,
while Appendix~\ref{app:A-source} gives an explicit second-order example.

\emph{Field dependence.} Each local cumulant $\kappa_m(t)$ is an integer
polynomial of degree at most $m$ and parity $(-1)^m$. Since the blocks
of $\pi$ contain $2n$ endpoints in total, the field factor in
Eq.~\eqref{eq:connected-partition-main} is even in $t$ and has degree
at most $2n$. Hence $\mathcal P_{n,L}$, and thus its thermodynamic
limit, is a polynomial in $y=t^2$ of degree at most $n$.
For $n=1$, Eq.~\eqref{eq:source-first-order} gives
$\mathcal P_1(y)=y/6$. For $n\ge2$, connectedness requires at least one
block of size $m\ge2$, and $\kappa_m(1)=0$; each term therefore
vanishes at $y=1$. This proves $\mathcal P_n(1)=0$ without cancellation
between graphs.

\emph{Spatial weights and rationality.} For each connected, loop-free
$G$, orient its edges and select a spanning tree. Inserting the discrete
Fourier transform of $A_L$ into Eq.~\eqref{eq:block-spatial-main}, the
site sums enforce momentum conservation at every vertex. The independent
momenta are the $r(G)=|E(G)|-|V(G)|+1$ fundamental cycle momenta, giving
\begin{equation}
 \mathcal A_{G,L}=\frac{1}{L^{r(G)}}
 \sum_{\bm k\in(\mathbb Z/L\mathbb Z)^{r(G)}}
 \prod_{e\in E(G)}\widehat A_L((C_G\bm k)_e).
 \label{eq:site-fourier-main}
\end{equation}
Here $C_G$ is the integral fundamental cycle matrix; its rows label
edges, and its columns label independent cycles. This formula uses the
unrestricted site sum, not a sum over distinct occupied sites. The
finite-size symbol in Eq.~\eqref{eq:A-symbol-main} converges to the
periodic Bernoulli function, so the normalized sum has the limit
\begin{equation}
 \mathcal A_G=\int_{[0,1]^{r(G)}}
 \prod_{e\in E(G)}2B_2\!\left(\{(C_G\bm p)_e\}\right)
 \,\dd^{r(G)}p,
 \label{eq:torus-graph-main}
\end{equation}
where $\{u\}=u-\lfloor u\rfloor$. The amplitude excludes the local
cumulants and the combinatorial weight in
Eq.~\eqref{eq:connected-partition-main}. For a tree, $r(G)=0$ and the
integral denotes evaluation at zero momentum; for a three-edge cycle,
\begin{equation}
 \mathcal A_{\triangle}=\int_0^1[2B_2(u)]^3\dd u=\frac{2}{945}.
 \label{eq:triangle-example}
\end{equation}
The hyperplanes $(C_G\bm p)_e\in\mathbb Z$ partition the unit cube
into rational polytopes on which the integrand is a polynomial with
rational coefficients. Its integral is therefore rational. Since
Eq.~\eqref{eq:connected-partition-main} is a finite sum with rational
combinatorial factors and integer-polynomial field factors, it follows
that $\mathcal P_n(y)\in\mathbb Q[y]$. Related methods for multivariable Bernoulli sums are discussed in \cite{Szenes1998,SzenesVergne2003}.
\end{proof}
The full field polynomial is useful even when one is ultimately interested in zero field.  Different derivatives with respect to the uniform source select different combinations of its coefficients.  For example, the linear and cubic magnetic responses at $h=0$ follow from
\begin{align}
 a_n^{(\chi)}&=[x^n]\left.\partial_h^2\phi(x,h)\right|_{h=0}
       =2\mathcal P_n'(0),\label{eq:AfromP-main}\\
 [x^n]\left.\partial_h^4\phi(x,h)\right|_{h=0}
       &=12\mathcal P_n''(0)-16\mathcal P_n'(0).\nonumber
\end{align}
Here primes on $\mathcal P_n$ denote derivatives with respect to $y=\tanh^2h$, and we used $\tanh^2h=h^2-\frac23h^4+O(h^6)$. Thus a single $\mathcal P_n$ contains, at order $x^n$, the coefficients of the linear susceptibility and all higher odd-order responses of the magnetization.

\subsection{Finite-ring structure and exact reconstruction}
For a finite system, one usually expects to approach the thermodynamic
limit by increasing the system size and extrapolating the resulting data.
The inverse-square chain has a remarkably stronger property: at each fixed
high-temperature order $n$, the dependence of the field-dependent free-energy
coefficient on ring length terminates \emph{exactly}. The finite-ring
structure is in fact an even polynomial in inverse length, with a sharp
degree bound that holds for each connected graph amplitude. This makes
finite-ring reconstruction algebraic rather than asymptotic.

\begin{theorem}[Sharp finite-size reconstruction]\label{thm:ring}
For every perturbative order $n\ge1$ and every integer ring length $L\ge2$,
\begin{equation}
 \mathcal P_{n,L}(y)=\sum_{j=0}^{n}c_{n,j}(y)L^{-2j},
 \qquad c_{n,j}(y)\in\mathbb Q[y],\qquad \deg_y c_{n,j}\le n.
 \label{eq:ring-poly-main}
\end{equation}
The thermodynamic coefficient is $\mathcal P_n(y)=c_{n,0}(y)$. Hence any
$n+1$ distinct integer ring lengths $L_0,\ldots,L_n\ge2$ determine it exactly:
\begin{equation}
 \mathcal P_n(y)=\sum_{a=0}^{n}\mathcal P_{n,L_a}(y)
 \frac{L_a^{2n}}{\displaystyle\prod_{b\ne a}(L_a^2-L_b^2)}.
 \label{eq:field-interp-main}
\end{equation}
\end{theorem}
For the linear magnetic response, define the reduced zero-field
susceptibility at fixed $x$ by
\begin{align}
 \chi_L(x)&=\left.\frac{\partial^2\phi_L(x,h)}{\partial h^2}\right|_{h=0}
 =1+\sum_{n\ge1}a_{n,L}^{(\chi)}x^n,\nonumber\\
 \chi(x)&=\left.\frac{\partial^2\phi(x,h)}{\partial h^2}\right|_{h=0}
 =1+\sum_{n\ge1}a_n^{(\chi)}x^n.
 \label{eq:chi-series-main}
\end{align}
The coefficient identities $a_{n,L}^{(\chi)}=2\mathcal P'_{n,L}(0)$
and $a_n^{(\chi)}=2\mathcal P'_n(0)$ then show that the same $n+1$
rings reconstruct every zero-field susceptibility coefficient:
\begin{equation}
 a_n^{(\chi)}=\sum_{a=0}^{n}a_{n,L_a}^{(\chi)}
 \frac{L_a^{2n}}{\displaystyle\prod_{b\ne a}(L_a^2-L_b^2)}.
 \label{eq:chi-interp-main}
\end{equation}
For the consecutive lengths $L=2,\ldots,n+2$ used in the calculations,
the weights take an explicit factorial form,
\begin{equation}
  a_n^{(\chi)}=\sum_{L=2}^{n+2}(-1)^{n+2-L}
 \frac{2(L^2-1)L^{2n+2}}{(n+2-L)!(n+2+L)!}
 a_{n,L}^{(\chi)}.
 \label{eq:chi-factorial-main}
\end{equation}
Thus the complete magnetic-field coefficient and its zero-field
susceptibility specialization are both exact finite combinations of
microscopic ring calculations.  The short formulas do not assume a
large-$L$ scaling form.

\noindent\emph{Proof outline.}
For a connected graph $G$ with $n$ edges and cycle rank $r$, the
finite-ring counterpart of Eq.~\eqref{eq:torus-graph-main} is the normalized
discrete momentum sum
\begin{equation}
 \mathcal A_{G,L}=\frac1{L^r}
 \sum_{\bm k\in(\mathbb Z/L\mathbb Z)^r}
 \prod_{e\in E(G)}\widehat A_L\bigl((C_G\bm k)_e\bigr),
 \qquad r=n-|V(G)|+1.
 \label{eq:finite-graph-main}
\end{equation}
Momentum conservation at the vertices fixes the bond momenta in terms
of independent cycle momenta, using integer coefficients. The special
structure of these graph constraints makes the polynomial pieces of the
Bernoulli Fourier symbol compatible with the discrete momentum grid.
The exact Fourier symbol, Eq.~\eqref{eq:A-symbol-main}, includes a
correction $-1/(3L^2)$. An exact count of the momentum-grid points then
shows that no additional dependence on whether $L$ is odd or even can
occur. In particular, no inverse power beyond $L^{-(r+2n)}$ can occur,
and $r+2n\le3n$; the detailed derivation
is given in Lemma~\ref{lem:finite-grid}.

The remaining steps sharpen this bound.  A second exact expression
for the same sine-square interaction is especially useful here.  At the
$L$th roots of unity $\zeta_j=e^{2\pi i j/L}$, and for unequal sites,
\begin{equation}
 L^2A_L(i-j)=4\mathscr K(\zeta_i,\zeta_j)\mathscr K(\zeta_j,\zeta_i),\qquad
 \mathscr K(a,b)=\frac{a}{a-b}.
 \label{eq:root-kernel-main}
\end{equation}
This form lets us sum the positions of the graph vertices one at a time.
The result is that $L^{2n}\mathcal A_{G,L}$ is an ordinary polynomial in
$L$, including contributions where different graph vertices occupy the
same site.  The details of the site sums are given in
Appendix~\ref{app:A-reconstruction}.

The Fourier description supplies the complementary observation: the
finite-grid sums contain only even powers of $L^{-1}$.  This follows
from the integer structure of the momentum grid and the continuity of
the periodic Bernoulli polynomial where its polynomial pieces meet.
The cancellation at intersections of these pieces is important;
Appendix~\ref{app:A-reconstruction} gives the full boundary argument.
Together, the two descriptions show that
$L^{2n}\mathcal A_{G,L}$ is an \emph{even} polynomial in $L$.
Finally, summing along a spanning tree bounds its growth by $O(L^{2n})$.
It follows that $\mathcal A_{G,L}$ contains no powers below $L^{-2n}$.
Since the connected-source factors are polynomials in $y$ independent of
$L$, the same conclusion holds for the complete field coefficient.

The complete proof, including repeated bonds, site coincidences, and
the independent conservative interpolation formula, is in
Appendix~\ref{app:A-reconstruction}.  Interpolation at $L^{-2}=0$
then gives Eq.~\eqref{eq:field-interp-main}; it determines the bulk
coefficient algebraically rather than by fitting a large-$L$ scaling
form.  At arbitrary order the practical limitation is the cost of the
finite-ring calculation.

\paragraph{The first two orders.}
The finite-ring formulas already illustrate the theorem.  Let
\begin{align}
 a_L&=\sum_{j\ne i}A_L(i-j)=\frac{1-L^{-2}}3,\nonumber\\
 b_L&=\sum_{j\ne i}A_L(i-j)^2
 =\frac{1+10L^{-2}-11L^{-4}}{45}.
 \label{eq:ab-ring-example}
\end{align}
The expression for $b_L$ follows from the discrete Fourier symbol by Parseval's identity.  The first coefficient is $\mathcal P_{1,L}(y)=ya_L/2$.  For the second, the variance of the interaction energy in the independent-spin measure yields
\begin{equation}
 \mathcal P_{2,L}(y)=\frac{1-y^2}{4}b_L
 +\frac{y(1-y)}2\bigl(a_L^2-b_L\bigr).
 \label{eq:P2-ring-example}
\end{equation}
The first term comes from two occurrences of the same edge; the second comes from pairs of distinct edges with one common endpoint.  Both expressions are polynomials in $L^{-2}$, and their constant terms give
$\mathcal P_1(y)=y/6$ and $\mathcal P_2(y)=(1-y)(1+9y)/180$.
For instance, at zero field $\mathcal P_{2,L}(0)=b_L/4$ takes the values $1/64$, $8/729$, and $9/1024$ for $L=2,3,4$, respectively.  The unique quadratic in $L^{-2}$ passing through these values has constant term $1/180$.  This makes the finite-interpolation statement concrete: no assumed scaling exponent or sequence extrapolation enters.

Theorem~\ref{thm:ring} now supplies a route from finite spin sums to exact thermodynamic field polynomials.  The next section shows what those polynomials look like and what the resulting coefficient array suggests.

\section{Explicit field coefficients and their triangular array}
Using the sharp finite-size reconstruction theorem, we have reconstructed the complete field-dependent free-energy polynomials $\mathcal P_1(y),\ldots,\mathcal P_{26}(y)$. Their coefficients form a triangular array,
\begin{equation}
 \mathcal P_n(y)=\sum_{m=0}^{n}p_{n,m}y^m,
 \qquad p_{n,m}\in\mathbb Q,
 \label{eq:coeff-triangle-definition}
\end{equation}
containing $\sum_{n=1}^{26}(n+1)=377$ rational entries, including the zero entry $p_{1,0}=0$. Thus the calculation gives the full magnetic-field dependence at each of the first 26 high-temperature orders, rather than just a single coefficient per order. The low powers of $y$ determine weak-field responses, while the highest-degree coefficients exhibit the exact all-order structure developed in Section~5.

The first four polynomials are
\begin{align}
\mathcal P_1(y)&=\frac{y}{6},\nonumber\\
\mathcal P_2(y)&=\frac{(1-y)(1+9y)}{180},\nonumber\\
\mathcal P_3(y)&=\frac{(1-y)(2+63y-135y^2)}{5670},\nonumber\\
\mathcal P_4(y)&=\frac{(1-y)(9+841y-4875y^2+4725y^3)}{340200}
 \label{eq:P1P4-main}
\end{align}
The further explicit expressions $\mathcal P_5,\ldots,\mathcal P_7$ are collected in Appendix~\ref{app:A-extra}.  For these low orders, the overall factor $1-y$ at $n\ge2$ and the finite degree are visible directly.  Each of the coefficients of $y,y^2,\ldots,y^n$ is a rational number, so its sign and its variation with order can be compared exactly.

To construct this array, magnetization-resolved energy moments were calculated exactly on periodized rings by enumerating rotational equivalence classes of spin configurations. For each order $n$, the reconstructed bulk row uses $L=2,\ldots,n+2$ and the sharp finite-size theorem. Eighteen primes near $2^{61}$ were used for rational reconstruction, and four unused primes checked the resulting values independently.

Across all archived ring lengths beyond the interpolation input, 73,008 additional modular comparisons passed. A separate full-spin implementation checked finite-ring series through order 26 on six smaller rings. Applying the independent conservative degree-$3n$ interpolation to the archived modular data also agrees with the reported rational values through order 8 on the 18 calculation primes and through order 9 on the four reserved primes. These last comparisons are exact modular checks, not an independent certificate of the bulk rational values without bounds on rational reconstruction. Details and executable verification materials are given in Appendix~\ref{app:A-reconstruction} and the Data Availability statement.

The explicit low-order polynomials through order seven are supplied in the text and Appendix~\ref{app:A-extra}.  The computational record for the higher orders is described in Appendix~\ref{app:A-reconstruction}; data availability is stated at the end of the paper.

The reconstructed coefficients through order 26 show that all 351 nonconstant coefficients obey
\begin{equation}
 \operatorname{sgn}\bigl([y^m]\mathcal P_n(y)\bigr)=(-1)^{m+1},\qquad 1\le m\le n\le26.
 \label{eq:sign-conj-main}
\end{equation}
The observed signs motivate the following all-order conjecture.
\begin{conjecture}[All-order alternating sign rule]\label{conj:sign}
For every $n\ge1$ and $1\le m\le n$, the coefficient $[y^m]\mathcal P_n(y)$ is nonzero and has sign $(-1)^{m+1}$.
\end{conjecture}

Figure~\ref{fig:triangle} illustrates Conjecture~\ref{conj:sign} for the first seven orders, using signs extracted from the computed rational coefficients.

\begin{figure}[htbp]
\centering
\begin{tikzpicture}[x=0.56cm,y=0.56cm]
  \foreach \m in {1,...,7} {\node[font=\scriptsize] at (\m,0.25) {\m};}
  \node[font=\scriptsize] at (4,1.05) {$m$ (field power)};
  \foreach \n in {1,...,7} {
    \node[font=\scriptsize,anchor=east] at (0.48,-\n) {\n};
  }
  % The following signs are read from the archived rational coefficients,
  % rather than generated from the conjectured parity rule.
  \foreach \n/\m/\signmark in {
      1/1/+, 2/1/+, 2/2/-, 3/1/+, 3/2/-, 3/3/+, 4/1/+,
      4/2/-, 4/3/+, 4/4/-, 5/1/+, 5/2/-, 5/3/+, 5/4/-,
      5/5/+, 6/1/+, 6/2/-, 6/3/+, 6/4/-, 6/5/+, 6/6/-,
      7/1/+, 7/2/-, 7/3/+, 7/4/-, 7/5/+, 7/6/-, 7/7/+
  } {
    \pgfmathtruncatemacro{\offset}{\n-\m}
    \ifnum\offset=0
      \node[circle,draw,inner sep=1.5pt,font=\scriptsize] at (\m,-\n) {$\signmark$};
    \else
      \ifnum\offset=1
        \node[rectangle,draw,inner sep=2pt,font=\scriptsize] at (\m,-\n) {$\signmark$};
      \else
        \node[font=\scriptsize] at (\m,-\n) {$\signmark$};
      \fi
    \fi
  }
  \node[rotate=90,font=\scriptsize] at (-0.3,-4) {$n$ (order)};
  \node[anchor=west,font=\footnotesize,text width=5.5cm] at (8,-2.6) {Circle: exact highest-degree coefficient $d_0(n)$.\\ Rectangle: exact first subleading coefficient $d_1(n)$.\\ Other entries: lower-degree coefficients.};
\end{tikzpicture}
\caption{Alternating-sign pattern of the field-polynomial coefficients for orders $n=1,\ldots,7$, illustrating Conjecture~\ref{conj:sign}. Signs alternate with the field power $m$, independently of the expansion order $n$. The signs shown are taken from the computed rational coefficients, not assigned from the conjecture. Circles and squares mark the highest- and next-to-highest-degree coefficients in $y$; their signs are established at every order by Theorems~\ref{thm:matching} and~\ref{thm:d1}, respectively.}
\label{fig:triangle}
\end{figure}

The pattern in Conjecture~\ref{conj:sign} is stronger than the polynomial degree bound: the latter restricts which powers of $y$ can occur but does not control their signs. The observed sign depends on the field power $m$, not on the perturbative order $n$. The reconstructed order-26 array supports this behavior; the finite-size reconstruction theorem does not itself impose a coefficientwise sign rule. Two parts of the sign rule, however, are proved at every order: Theorem~\ref{thm:matching} establishes the sign of $[y^n]\mathcal P_n$, and Theorem~\ref{thm:d1} establishes the sign of $[y^{n-1}]\mathcal P_n$ for $n\ge2$. We turn next to these two exact sequences.

\section{Closed-form leading field coefficients and the vertex-collision expansion}
The coefficients with the highest powers of $y=\tanh^2h$ have particularly simple forms. Instead of calculating $\mathcal P_n$ in full, we first organize the normalized partition function by the number of interaction-bond endpoints occupying the same site. At order $x^n$, the highest power $t^{2n}$, with $t=\tanh h$, comes from $n$ disjoint bonds, forming a matching. Taking the logarithm then produces the connected free-energy contribution: in a direct cumulant expansion, repeated bonds and shared sites can still contribute to $[y^n]\mathcal P_n$. The matching method remains exact because extracting the highest allowed field power commutes with the formal logarithm, as explained below and in Appendix~\ref{app:A-matching}. Lower field powers are organized by successive local collisions before taking the logarithm. The highest-degree coefficient by itself does not describe the strong-field limit: as $y\to1$, different powers cancel to give $\mathcal P_n(1)=0$ for $n\ge2$.

We label the coefficients by their distance below the highest power of $y$ (not by their large-field asymptotics):
\begin{equation}
 d_k(n)=[y^{n-k}]\mathcal P_n(y).
\end{equation}
The index $k$ has a simple combinatorial meaning in the spin expansion.  For a multigraph $G$ with $n$ interaction edges, let $o(G)$ denote its number of odd-degree vertices.  In the independent-spin measure the spin product associated with $G$ has expectation $t^{o(G)}$, with $t=\tanh h$.  Thus
\begin{equation}
 k(G)=n-\frac{o(G)}2
     =\sum_v\left\lfloor\frac{d_v}{2}\right\rfloor.
 \label{eq:defect-grade-main}
\end{equation}
The integer $k(G)$ counts the number of pairs of interaction-bond endpoints merged at vertices, with multiplicity; we call it the \emph{vertex-collision index}.  These combinatorial collisions are \emph{not} physical domain walls or kinks.  For $k(G)=0$, all $2n$ endpoints are distinct and the bonds form a matching.  For $k(G)=1$, one vertex has degree two or three, while the remaining bonds form a matching.  The logarithm of the partition function selects the connected contributions.  This organization by decreasing powers of $y$ leads to the determinant constructions below; the full representation is given in Appendix~\ref{app:A-defect}.

\begin{theorem}[Highest-degree field coefficient from a matching determinant]\label{thm:matching}
For every $n\ge1$, the highest-degree coefficient of the field polynomial is given exactly by
\begin{equation}
 d_0(n)=\frac{(-1)^{n+1}}{2n(2n+1)}.
 \label{eq:d0-main}
\end{equation}
\end{theorem}
The coefficient of $x^nt^{2n}$ in the normalized partition function is a weighted sum over matchings of $n$ bonds \cite{HeilmannLieb1972}. Because maximal field degrees add under multiplication, selecting this coefficient commutes with taking the formal logarithm. If $d_{0,L}(n)=[y^n]\mathcal P_{n,L}(y)$ and $\mathfrak M_L(w)$ denotes the finite-ring matching polynomial, this gives the exact identity
\begin{equation}
 \sum_{n\ge1}d_{0,L}(n)w^n=\frac1L\log\mathfrak M_L(w).
 \label{eq:matching-log-main}
\end{equation}
The logarithm supplies the connected subtractions; it is not restricted to disjoint-bond graphs when expanded directly in cumulants. The matching weights use the periodized interaction $A_L$ defined in Section~\ref{sec:tools}. Introduce the skew-symmetric matrix
\begin{equation}
 (K_L)_{ij}=\begin{cases}
 \displaystyle\frac1L\csc\!\frac{\pi(i-j)}{L},&i\ne j,\\[2mm]
 0,&i=j.
 \end{cases}
 \label{eq:K-def-main}
\end{equation}
Its entrywise square equals $A_L(i-j)$ off the diagonal.  The weighted matching polynomial is a Cauchy determinant:
\begin{equation}
 \mathfrak M_L(w)=\det(I+\sqrt w\,K_L)
 =\prod_{k=0}^{\lfloor L/2\rfloor-1}
 \left[1+w\left(\frac{L-1-2k}{L}\right)^2\right].
 \label{eq:matching-main}
\end{equation}
Substituting this determinant into Eq.~\eqref{eq:matching-log-main} gives the highest-degree free-energy coefficient at every order.  Taking the large-$L$ limit gives
\begin{equation}
 \sum_{n\ge1}d_0(n)w^n
 =\frac14\int_{-1}^1\log(1+wu^2)\,\dd u,
 \label{eq:matching-thermo-main}
\end{equation}
which immediately yields the formula in Theorem~\ref{thm:matching} by expanding the logarithm.  The determinant--Hafnian identity and skew-kernel spectrum are derived in Appendix~\ref{app:A-matching}; related determinant identities are given in \cite{IshikawaEtAl2005}.

\begin{theorem}[First subleading field coefficient]\label{thm:d1}
For every $n\ge2$, the next-to-highest-degree coefficient in $y$ is exactly
\begin{equation}
 d_1(n)=(-1)^n\frac{5n-4}{9(2n-1)(2n+1)},\qquad n\ge2.
 \label{eq:d1-main}
\end{equation}
\end{theorem}
At vertex-collision index one, a single exceptional site has degree two or three, with every other occupied site of degree one. Fixing that site, the remaining disjoint bonds form a matching on the undeleted sites. Each choice of two or three leaves therefore gives a deleted principal minor of the matching matrix. Jacobi's complementary-minor identity turns the sum of such minors into coefficients of one determinant quotient. A diagonal conjugation then identifies the quotient with the resolvent of a finite cycle Laplacian, whose spectrum is elementary. Appendix~\ref{app:A-d1} gives the determinant, the cycle matrix elements, and an explicit cyclic-gap sum before carrying out the thermodynamic limit. The formula agrees with the corresponding coefficients of $\mathcal P_2$ and $\mathcal P_3$, as well as with the higher-order reconstructed values.

It is natural to ask whether further subleading sequences admit similar expressions.  Their subgraphs with shared bond endpoints become more numerous, and products of contributions at smaller vertex-collision index must be subtracted when taking the connected logarithm.  The source determinant at fixed vertex-collision index keeps this combinatorics finite; the further simplification to rational functions of $n$ is presently supported by reconstruction and independent tests on rows not used to determine those functions.

\begin{conjecture}[Second subleading coefficient: rational formula]\label{conj:d2}
The reconstructed coefficients through order 26 agree with the following expression, conjectured to hold for $n\ge3$:
\begin{equation}
 d_2(n)=(-1)^{n+1}\frac{1250n^3-5132n^2+6534n-2589}
 {4050(2n-3)(2n-1)(2n+1)},\qquad n\ge3,
 \label{eq:d2-main}
\end{equation}
with $d_2(2)=1/180$.
\end{conjecture}
Once the displayed denominator is assumed, four coefficients, at $n=3,4,5,6$, determine its cubic numerator; the remaining 20 rows $n=7,\ldots,26$ agree exactly. The equivalent seven-parameter generating function also reproduces those rows, but this is a different form of the same proposed sequence, not an independent test. The missing all-order step is the simplification of the connected shared-endpoint contributions in the source determinant, as discussed in Appendix~\ref{app:A-d2}.

Appendix~\ref{app:A-extra} gives the third subleading sequence $d_3(n)$ explicitly.  For offsets $k=1,\ldots,8$, determining a proposed numerator polynomial from the minimum number of reconstructed rows leaves respectively $23,20,17,14,11,8,5,2$ further rows for checking. All agree with the following pattern, although the evidence becomes less restrictive as $k$ increases: at $k=8$, only two rows remain beyond the fit.
\begin{conjecture}[Rational formulas at fixed field-degree offset]\label{conj:all-k}
For each fixed $k\ge1$, $d_k(n)$ has the rational form
\begin{equation}
 (-1)^{n+k+1}d_k(n)\prod_{j=0}^{k}(2n+1-2j)=R_k(n),
 \qquad \deg R_k=2k-1.
 \label{eq:dk-pattern-main}
\end{equation}
Here $R_k$ is a polynomial with rational coefficients. The conjectured range is $n\ge k+1$; the boundary coefficient $d_k(k)=[y^0]\mathcal P_k(y)$ is not included and must be treated separately.
\end{conjecture}
The cases $k=0$ and $k=1$ are established by Theorems~\ref{thm:matching} and \ref{thm:d1}; for $k\ge2$, the denominator factors and the degree $2k-1$ are empirical regularities, not consequences of the matching determinant. The checks through $k=8$ support the proposed range $n\ge k+1$. In particular, fixed $k$ describes a narrow set of coefficients near the highest field degree, rather than the zero-field susceptibility at arbitrary order.

The vertex-collision index in Eq.~\eqref{eq:defect-grade-main} also organizes the source construction. At fixed $k$, at most $k$ sites are exceptional; all remaining occupied sites form a matching background. Appendix~\ref{app:A-defect} reduces their contribution to finitely many derivatives of one determinant. The rational-form conjecture also restricts the corresponding generating functions. Define $G_k(w)=\sum_{n\ge k}d_k(n)w^n$. The conjectured rational form implies
\begin{equation}
 \begin{aligned}
 G_k(w)={}&U_k(w)\frac{\arctan\sqrt w}{\sqrt w}
 +\sum_{r=1}^{k-1}\frac{b_{k,r}}{(1+w)^r}\\
 &+V_k(w),
 \end{aligned}
 \label{eq:Vk-main}
\end{equation}
where $U_k$ and $V_k$ are polynomials of degree at most $k$ with rational coefficients, and the $b_{k,r}$ are rational constants. To see the implication, divide $R_k(n)$ by the denominator in Eq.~\eqref{eq:dk-pattern-main} and use partial fractions. The polynomial part has degree at most $k-2$, giving poles at $w=-1$ of order at most $k-1$; the remaining simple fractions sum to shifts of $\arctan\sqrt w/\sqrt w$, plus finite polynomials. The finitely many coefficients outside $n\ge k+1$ affect only the polynomial term. Thus Eq.~\eqref{eq:Vk-main} is a consequence of the rational-form conjecture, not independent evidence for it.

The relation $a_n^{(\chi)}=2d_{n-1}(n)$ shows that the zero-field susceptibility probes coefficients whose vertex-collision index grows with $n$. Thus the fixed-$k$ formulas complement, rather than replace, the usual weak-field series. They also identify a concrete open task: derive the proposed fixed-offset rational formulas directly from the connected source determinant. We now turn from the algebraic structure of the coefficients to the magnetic observables they determine.

\section{Magnetic response and continuation to imaginary field}
The 26 field polynomials obtained above contain more information than the free-energy coefficients at zero field. At each high-temperature order, derivatives of the same polynomial generate the magnetization and the hierarchy of nonlinear magnetic responses, without a separate series calculation for each observable. We illustrate this direct access to real-field response and then continue the expansion to imaginary magnetic field, where the coefficient structure can be examined in the setting of Lee--Yang theory \cite{LeeYang1952,YangLee1952}.

The reduced field is $h=\beta H_{\mathrm{ext}}$, and $m=\partial_h\phi$ is the magnetization per spin, measured in units of the spin magnetic moment.  Differentiating $\phi=-\beta f$ gives
\begin{equation}
 m(x,h)=t+2t(1-t^2)\sum_{n\ge1}\mathcal P_n'(t^2)x^n,
 \qquad t=\tanh h.
 \label{eq:eos-main}
\end{equation}
A second derivative with respect to $h$ gives the reduced uniform susceptibility $\chi_h=\partial_hm$.  At fixed temperature the response to the applied field energy is $\chi_{H}=\partial m/\partial H_{\mathrm{ext}}=\beta\chi_h$.  If $H_{\mathrm{ext}}=\mu B$ for a laboratory field $B$ and spin magnetic moment $\mu$, the physical susceptibility of the magnetic moment per site is $\chi_B=\partial(\mu m)/\partial B=\beta\mu^2\chi_h$.  Writing $S_1=\sum_n\mathcal P_n'(y)x^n$ and $S_2=\sum_n\mathcal P_n''(y)x^n$, one obtains
\begin{equation}
 \chi_h=(1-y)\left[1+2(1-3y)S_1+4y(1-y)S_2\right].
\end{equation}
The calculated polynomials therefore give $m$ and $\chi_h$ through order $x^{26}$ as explicit functions of $h$, and further field derivatives yield higher nonlinear responses to the same order. As one illustration, consider the stationary points of the truncated susceptibility as a function of $x$ at fixed nonzero reduced field $h$. Since $h=\beta H_{\mathrm{ext}}$, these are fixed-reduced-field characteristics of the series rather than susceptibility peaks at fixed applied field. For a truncation at order $N$, the stationary-point equation is
\begin{equation}
 (1-3y)\sum_{n=1}^{N}n\mathcal P_n'(y)x^{n-1}
 +2y(1-y)\sum_{n=1}^{N}n\mathcal P_n''(y)x^{n-1}=0.
 \label{eq:crossover-main}
\end{equation}
For the order-26 polynomials, the smallest positive stationary roots are listed in Table~\ref{tab:chi-stationary}.

\begin{table}[!htbp]
\centering
\caption{Stationary points of the order-26 truncated reduced susceptibility as a function of coupling at fixed reduced field $h$. Here $t=\tanh h$, $x_*$ is the smallest positive solution of Eq.~\eqref{eq:crossover-main}, and $g_*=x_*/\pi^2$.}
\label{tab:chi-stationary}
\begin{tabular}{c c c c}
\toprule
$t=\tanh h$ & $h$ & $x_*(h)$ & $g_*(h)=x_*/\pi^2$\\
\midrule
$1/4$&0.25541&2.07667&0.21041\\
$1/5$&0.20273&2.44162&0.24739\\
$1/8$&0.12566&3.06733&0.31079\\
$1/20$&0.05004&3.88021&0.39315\\
$1/25$&0.04002&3.99276&0.40455\\
$1/32$&0.03126&4.14973&0.42046\\
$1/40$&0.02501&4.31499&0.43720\\
\bottomrule
\end{tabular}
\end{table}

Over the range shown in Table~\ref{tab:chi-stationary}, the stationary point shifts to larger coupling as the reduced field decreases. This illustrates the field dependence of the order-26 magnetic response.

\subsection{Analytic continuation to imaginary magnetic field}
Retaining the magnetic field also allows the same high-temperature coefficients to be continued directly to imaginary $h$, without repeating the lattice calculation. In this domain, the partition-function zeros lie on the Lee--Yang circle, providing a complementary perspective on magnetic-field analyticity \cite{LeeYang1952,GlumacUzelac1991}. The continuation also reveals a direct consequence of the alternating-sign pattern observed in Section~4.

For $h=i\theta$ with real $\theta$, set $q=\tan\theta$. The magnetization is imaginary, so for $q\ne0$ it is convenient to factor out its leading value $iq$ and define
\begin{equation}
 M_q(x)=\frac{m(x,i\theta)}{iq}
 =1+2(1+q^2)\sum_{n\ge1}\mathcal P_n'(-q^2)x^n.
 \label{eq:Mq-main}
\end{equation}
At $q=0$ we use the continuous limit $M_0(x)=\lim_{q\to0}M_q(x)=\chi_h(x,0)$; in particular, the zero-field ratio estimates below use this definition. The alternating-sign pattern of Conjecture~\ref{conj:sign} has a simple consequence: each term $m p_{n,m}(-q^2)^{m-1}$ in $\mathcal P_n'(-q^2)$ is positive when the sign rule holds. Accordingly, every nonconstant coefficient of $M_q(x)$ obtained through order 26 is positive for $q>0$. The sign structure of the field polynomials thus becomes a positivity property of the computed imaginary-field coupling series.

To relate this continuation to the Lee--Yang description, introduce the field fugacity $z=e^{-2h}$. For $h=i\theta$, one has $z=e^{-2i\theta}$: the positive zero angle $\varphi$ corresponds to $\theta=-\varphi/2$ modulo $\pi$.  For conjugate Lee--Yang zeros $z_j=e^{\pm i\varphi_j}$, $0<\varphi_j\le\pi$, define $g_{\rm LY}(\varphi)$ as the thermodynamic zero density per spin on the \emph{positive-angle half} of the unit circle, normalized by $\int_0^\pi g_{\rm LY}(\varphi)\,\dd\varphi=1/2$.  With this convention the magnetization admits the representation
\begin{equation}
 m(h)=2\sinh2h\int_{\varphi_e}^{\pi}\frac{g_{\rm LY}(\varphi)}{\cosh2h-\cos\varphi}\,\dd\varphi.
 \label{eq:LY-density-main}
\end{equation}
If the Lee--Yang edge remains at $\varphi_e>0$, then $m(h)\to0$ as $h\downarrow0$.  If the edge closes and the positive-angle zero density has a finite right limit at the origin,
\begin{equation}
 m_+=2\pi g_{\rm LY}(0^+).
 \label{eq:jump-LY-main}
\end{equation}
Here $m_+$ is the positive-field spontaneous magnetization. By spin-reversal symmetry the negative-field limit is $m_-=-m_+$, so the full magnetization discontinuity is $\Delta m=m_+-m_-=2m_+=4\pi g_{\rm LY}(0^+)$. Edge closure alone does not determine the size of this jump.

For $M_q(x)=1+\sum_nc_n(q)x^n$, define
\begin{equation}
 r_n(q)=\frac{c_{n-1}(q)}{c_n(q)},\qquad
 \rho_n(q)=nr_n(q)-(n-1)r_{n-1}(q).
\end{equation}
For a dominant algebraic singularity with a regular large-order expansion, this combination cancels the leading $1/n$ correction to the ratios and gives $\rho_n(q)=x_{\mathrm{sing}}(q)+O(n^{-2})$, where $x_{\mathrm{sing}}(q)$ is the leading singularity in the \emph{coupling} variable at fixed imaginary field.  The behavior is altered by competing singularities or confluent corrections.  Table~\ref{tab:imag-field-ratios} gives the finite-order coupling-series ratio diagnostics obtained from the reconstructed field polynomials. At moderate $q$ the estimates from successive orders are close, while the spread increases toward smaller imaginary fields.

\begin{table}[!htbp]
\centering
\caption{Imaginary-field coupling-series ratio diagnostics $\rho_n(q)$ from the order-26 field polynomials. The field is $h=i\theta$ with $q=\tan\theta$; the final column expresses $\rho_{26}$ in units of $g=x/\pi^2$.}
\label{tab:imag-field-ratios}
\small
\begin{tabular}{c r r r r}
\toprule
$q$&$\rho_{24}$&$\rho_{25}$&$\rho_{26}$&$\rho_{26}/\pi^2$\\
\midrule
$1/2$&1.456778&1.456936&1.457074&0.147633\\
$1/3$&1.988231&1.988478&1.988694&0.201497\\
$1/4$&2.356587&2.356911&2.357195&0.238834\\
$1/5$&2.628383&2.628776&2.629121&0.266386\\
$1/6$&2.839091&2.839549&2.839948&0.287747\\
$1/8$&3.148766&3.149390&3.149923&0.319154\\
$1/10$&3.369312&3.370011&3.370688&0.341522\\
$1/12$&3.537793&3.538034&3.538551&0.358530\\
$1/16$&3.788000&3.785807&3.784349&0.383435\\
$1/20$&3.971019&3.968113&3.964735&0.401712\\
\bottomrule
\end{tabular}
\end{table}

At $q=1/32$ the three values are $4.26576$, $4.28573$, and $4.29786$, whereas $\rho_{26}(0)=6.10956$. This growing spread illustrates the less stable ratio behavior near zero imaginary field.

\paragraph{Why a separate zero-field determinant fails.}
A sine-product expression reproduces the \emph{interaction part} $\phi(x,0)-\log2$ of the zero-field series through fifth order but fails at sixth order, when further connected even-degree graphs contribute. Appendix~\ref{app:A-sixth} explains this failure from the usual high-temperature graph expansion.

\section{Discussion and conclusion}\label{sec:discussion}
The inverse-square Ising chain has long provided a benchmark for unconventional ordering in one dimension, while its full field-dependent lattice thermodynamics remains unknown in closed form. This work establishes that its high-temperature expansion nevertheless has a remarkably rigid algebraic structure. At every perturbative order, the free-energy coefficient is a rational polynomial in $y=\tanh^2 h$ after rescaling the coupling by $\pi^2$. Moreover, its counterpart on a periodized ring is an exact polynomial in $L^{-2}$, of degree at most $n$ at order $n$. The constant term is the infinite-chain coefficient, so calculations on $n+1$ finite rings determine it exactly, with no numerical thermodynamic-limit extrapolation. This gives an in-principle reconstruction method at arbitrary high-temperature order.

The complete field dependence has been reconstructed through order 26, giving 26 polynomials with 377 rational coefficient entries, rather than one scalar coefficient per order as in earlier zero-field series \cite{Matvienko1985}. The sharp finite-size theorem makes $n+1$ ring sizes sufficient at order $n$. Independent spin enumeration, reserved primes, and additional-ring modular checks support the reported coefficients; these computational checks are distinct from the proof of the all-order interpolation theorem.

A second set of all-order results comes from the field-polynomial coefficients themselves. Matching determinants and a cycle-Laplacian reduction yield exact formulas for the two highest field-degree coefficients at every order. The full coefficient array also exhibits an alternating-sign pattern in its nonconstant entries and simple rational forms along several further diagonals. These observations suggest that the algebraic organization of the expansion extends well beyond the sequences presently accessible to proof.

The field polynomials also supply the magnetic equation of state and the hierarchy of nonlinear susceptibilities, order by order, without separate expansions for each observable. Section~6 illustrates this information through the real-field susceptibility and its continuation to imaginary magnetic field. In the latter case, the observed sign pattern implies positivity of the computed coupling-series coefficients, while finite-order coefficient ratios characterize the coupling expansion at fixed imaginary field. These applications show how the same rational coefficient array serves both thermodynamic response and questions of complex-field analyticity.

The main outcome is therefore not only a high-order series for a difficult long-range model, but an exact connection between finite-ring calculations and the infinite-chain high-temperature coefficients, complemented by analytical formulas for entire families within the field expansion. The combination of connected magnetic sources, the Bernoulli Fourier kernel, and determinant methods provides a framework for extending this microscopic information to higher orders and for investigating the further structures suggested by the 377 rational entries. The zero-field susceptibility coefficients, given by $a_n^{(\chi)}=2\mathcal P_n'(0)$, can be developed to higher orders without computing the complete field-polynomial array. A companion study uses these coefficients to investigate nested logarithmic susceptibility ratios, their conditional implications for critical parameters, and the critical turnover problem.
\clearpage
\appendix
\phantomsection
\addcontentsline{toc}{section}{Appendices}
\noindent{\Large\bfseries Appendices}\par
\vspace{\baselineskip}
% Print "Appendix A", "Appendix B", etc. in the appendix headings, without
% changing \thesection, references, equation numbers, or theorems.
\makeatletter
\let\original@sectionformat\@seccntformat
\renewcommand{\@seccntformat}[1]{%
  \ifnum\pdfstrcmp{#1}{section}=0
    Appendix~\thesection\quad
  \else
    \original@sectionformat{#1}%
  \fi}
\makeatother
\section{Detailed connected-source hierarchy}\label{app:A-source}
The source equation in Section~\ref{sec:tools} is exact for independent site fields; the key point for subsequent graph expansions is how it handles repeated interaction bonds and shared sites. We give an explicit second-order calculation and collect the local-cumulant properties needed in Section~3; connected expansions of this type are standard in lattice statistical mechanics \cite{FarrellMoritaMeijer1966}.

\subsection{The source identity and its connected coefficients}
At fixed $L$, differentiating the normalized partition function in Eq.~\eqref{eq:finite-source-main} gives
$\partial_x\cZ_L=\frac12\sum_{ij}A_{ij}\partial_i\partial_j\cZ_L$.
Using $\cZ_L=e^{W_L}$ and the initial condition
$W_L(0,\bm h)=\sum_i\log\cosh h_i$ yields the nonlinear equation
\eqref{eq:W-main} and its coefficient recursion \eqref{eq:W-rec-main}, which we use in the equivalent notation
\begin{equation}
 (n+1)W_{n+1,L}=\frac12\sum_{ij}A_{ij}
 \left[\partial_i\partial_jW_{n,L}
 +\sum_{r=0}^{n}(\partial_iW_{r,L})(\partial_jW_{n-r,L})\right].
 \label{eq:Wrecursion}
\end{equation}
The logarithm removes disconnected bond contributions; the local sources retain the information about which endpoints belong to the same site. The two terms on the right can act on different patterns of source derivatives, so multiple local cumulant blocks can occur at one site.

\subsection{Second-order example: repeated bonds and shared sites}
Set $t_i=\tanh h_i$ and let $\langle\cdot\rangle_{0,\bm h}$ denote expectation in the independent-spin measure at zero coupling but with the specified local fields, so that $\langle\sigma_i\rangle_{0,\bm h}=t_i$. With
\begin{equation}
 T=\sum_{i<j}A_{ij}\sigma_i\sigma_j,
 \qquad W_L(x,\bm h)-W_L(0,\bm h)
       =\log\langle e^{xT}\rangle_{0,\bm h},
 \label{eq:source-T}
\end{equation}
the first two coefficients are $W_{1,L}=\langle T\rangle_{0,\bm h}$ and
$W_{2,L}=\frac12[\langle T^2\rangle_{0,\bm h}-\langle T\rangle_{0,\bm h}^2]$.
Disjoint bonds have zero covariance. For a repeated bond and for two distinct bonds meeting at $i$, respectively,
\begin{align}
 \operatorname{Var}_{0,\bm h}(\sigma_i\sigma_j)&=1-t_i^2t_j^2,\nonumber\\
 \operatorname{Cov}_{0,\bm h}(\sigma_i\sigma_j,\sigma_i\sigma_k)
 &=t_jt_k(1-t_i^2),\qquad j\ne k.\nonumber
\end{align}
Keeping the correct factors from the expansion of $T^2$ gives
\begin{equation}
\begin{aligned}
 W_{2,L}={}&\frac12\sum_{i<j}A_{ij}^{2}(1-t_i^2t_j^2)\\
 &+\sum_i\sum_{\substack{j<k\\j,k\ne i}}
 A_{ij}A_{ik}\,t_jt_k(1-t_i^2).
\end{aligned}
\label{eq:source-W2}
\end{equation}
This expression is an identity at arbitrary local fields. On setting $t_i=t$ and $y=t^2$, translation invariance with
$a_L=\sum_{j\ne i}A_{ij}$ and $b_L=\sum_{j\ne i}A_{ij}^2$ gives
\begin{equation}
 \begin{aligned}
 \mathcal P_{2,L}(y)
 &=\frac{b_L}{4}(1-y^2)
   +\frac{a_L^2-b_L}{2}\,y(1-y)\\
 &=\frac{b_L}{4}(1-y)^2
   +\frac{a_L^2}{2}\,y(1-y).
 \end{aligned}
 \label{eq:source-P2}
\end{equation}
The first line separates repeated bonds from distinct bonds sharing a site, as in Eq.~\eqref{eq:P2-ring-example}. The second groups terms according to local cumulant blocks: parallel-edge and two-edge-path contributions, respectively. The path sum includes the possibility that its two leaves carry the same physical site label; this is why it contains $a_L^2$ rather than $a_L^2-b_L$.

For example, a repeated bond cannot be represented solely by one second cumulant at each endpoint, because
\begin{equation}
 \begin{aligned}
 1-t_i^2t_j^2
 ={}&\kappa_2(t_i)\kappa_2(t_j)\\
 &+\kappa_2(t_i)t_j^2+t_i^2\kappa_2(t_j),
 \qquad \kappa_2(t)=1-t^2.
 \end{aligned}
 \label{eq:repeated-bond-blocks}
\end{equation}
Thus the endpoint partitions must be retained when passing from the exact source recursion to the graph expansion. The finite-size formulas for $a_L$ and $b_L$ in Eq.~\eqref{eq:ab-ring-example} then give
$\mathcal P_2(y)=(1-y)(1+9y)/180$, agreeing with the independent finite-ring calculation.

\subsection{Local spin cumulants}
For a uniform field write
\begin{equation}
 t=\tanh h,
 \qquad
 \kappa_m(t)=\frac{\dd^m}{\dd h^m}\log\cosh h,
 \qquad m\ge1.
 \label{eq:kappa-field}
\end{equation}
Since $\dd/\dd h=(1-t^2)\dd/\dd t$,
\begin{equation}
 \kappa_1(t)=t,
 \qquad
 \kappa_{m+1}(t)=(1-t^2)\frac{\dd}{\dd t}\kappa_m(t).
 \label{eq:kappa-rec}
\end{equation}
In particular $\kappa_m\in\mathbb Z[t]$, $\deg\kappa_m=m$, and
\begin{equation}
 \kappa_m(-t)=(-1)^m\kappa_m(t).
 \label{eq:kappa-parity}
\end{equation}
For $m\ge2$, $\kappa_m$ is divisible by $1-t^2$. At zero field the odd cumulants vanish, while the even ones are integers:
\begin{equation}
 u_{2m}=\kappa_{2m}(0)
 =\frac{2^{2m}(2^{2m}-1)}{2m}B_{2m},
 \label{eq:local-cumulants}
\end{equation}
so $u_2=1$, $u_4=-2$, $u_6=16$, $u_8=-272$, and $u_{10}=7936$.
Together with the connected endpoint partitions described above, the parity and degree properties bound each order-$n$ field contribution by degree $2n$ in $t$, independently of the spatial weights. Those weights are treated separately in Appendix~\ref{app:A-reconstruction}.

\section{Rational field coefficients and exact finite-size reconstruction}\label{app:A-reconstruction}
We first prove rationality of the field-dependent coefficients using connected graph amplitudes. Related exact multiple-sum methods can be found in \cite{Mordell1958,LiuXin2026}, although the argument below needs only the piecewise-polynomial form of the interaction spectrum. We then prove the sharp finite-size reconstruction theorem by combining an independent finite-grid bound, toric weighted-Ehrhart parity, root-of-unity elimination, and a connected-graph degree estimate.  The independent matching-sector calculation is presented in Appendix~\ref{app:A-matching}; the remaining appendices develop its vertex-collision expansion.

\label{sec:rationality}

The construction in Section~\ref{sec:tools} has two parts: the local
sources generate the connected coefficients, while the exact Fourier
symbol evaluates their spatial sums. We supply the details behind
Eqs.~\eqref{eq:connected-partition-main}--\eqref{eq:torus-graph-main}
here, including the normalization and the role of coincident site labels.

\medskip
\noindent\textbf{Connected endpoint partitions.}
At fixed $L$ and arbitrary local sources, use the interaction variable
$T$ defined in Eq.~\eqref{eq:source-T}. The cumulant expansion reads
\begin{equation}
 W_L(x,\bm h)-W_L(0,\bm h)
 =\sum_{n\ge1}\frac{x^n}{n!}
   \operatorname{cum}_{0,\bm h}(T,\ldots,T).
 \label{eq:log-cumulant-all-orders}
\end{equation}
Replace $T$ by $\frac12\sum_{i,j}A_{ij}\sigma_i\sigma_j$ in each of
the $n$ arguments and regard its $2n$ spin factors as individually
labelled endpoints. Write
$\eta=\{\{1,2\},\ldots,\{2n-1,2n\}\}$ for the partition into the
original pairs. For fixed bond endpoints $(i_a,j_a)$, the cumulant-of-products
identity \cite{LeonovShiryaev1959} is
\begin{equation}
 \begin{aligned}
 &\operatorname{cum}_{0,\bm h}
       (\sigma_{i_1}\sigma_{j_1},\ldots,
        \sigma_{i_n}\sigma_{j_n})\\
 &\quad=\sum_{\substack{\pi\in\Pi(2n)\,\text{site-respecting}\\
                         \pi\vee\eta=\mathbf1}}
       \prod_{B\in\pi}\kappa_{|B|}(t_{s(B)}).
 \end{aligned}
 \label{eq:product-cumulant-graph}
\end{equation}
Here $\pi$ is \emph{site-respecting} for fixed endpoint labels if all
spin factors in any block $B$ are at one physical site $s(B)$; the join
$\pi\vee\eta=\mathbf1$ means that the blocks connect all the bond
occurrences. To see this identity, first expand every moment of the
bond observables into cumulants of individual spin factors. Applying
the usual moment--cumulant inversion to the outer joint cumulant
cancels exactly those endpoint partitions that do not connect the
bond groups. Independence at $x=0$ then eliminates blocks containing
different physical sites.

For uniform $h_i=h$, sum the original endpoint labels before fixing
$\pi$. Each block becomes a vertex of the connected multigraph
$G_\pi$, each pair of $\eta$ becomes an edge, and its site label is
summed independently. Thus \emph{different blocks may still have equal
site labels}; only an edge with coincident endpoints vanishes because
$A_{ii}=0$. The ordered-bond representation supplies $2^{-n}$, the
cumulant expansion supplies $1/n!$, and translation reduction supplies
$1/L$. This gives precisely Eq.~\eqref{eq:connected-partition-main}
with the spatial sum in Eq.~\eqref{eq:block-spatial-main}. For example,
its order-two contribution reproduces both decompositions in
Eq.~\eqref{eq:source-P2}. Requiring distinct labels for separate blocks
would instead give different graph amplitudes and would incorrectly
remove some repeated-bond contributions.

\medskip
\noindent\textbf{Fourier reduction of the unrestricted spatial sum.}
Let $G=(V,E)$ be a connected, loop-free cumulant-block multigraph with
$n=|E|$. After orienting each edge $e=(u,v)$, use
\begin{equation}
 A_L(s_u-s_v)=\frac1L\sum_{k_e\in\mathbb Z/L\mathbb Z}
 \widehat A_L(k_e)
 e^{2\pi i k_e(s_u-s_v)/L}.
 \label{eq:appendix-edge-fourier}
\end{equation}
Let $D_G$ be its oriented vertex--edge incidence matrix. Summing each
site label in Eq.~\eqref{eq:block-spatial-main} produces the constraints
$D_G\bm k=0\pmod L$. A spanning tree gives an integral basis for
these solutions: there are $r=n-|V|+1$ freely chosen cycle momenta,
and $\bm k=C_G\bm q$, where $C_G$ is the fundamental cycle matrix.
This parametrization is valid for every integer $L$, since the reduced
incidence matrix of a spanning tree is unimodular. The $|V|$ site sums
contribute $L^{|V|}$, which cancel the $L^{n+1}$ from the inverse
Fourier transforms and the prefactor in Eq.~\eqref{eq:block-spatial-main}
up to $L^{-r}$. Thus the result is exactly
Eq.~\eqref{eq:site-fourier-main}, including its normalization.

The finite-$L$ symbol in Eq.~\eqref{eq:A-symbol-main} is a continuous
periodic Bernoulli function sampled at the lattice momenta, up to its
explicit $-1/(3L^2)$ correction. At fixed $G$, the normalized finite
sum therefore converges to the torus integral
\begin{equation}
 \mathcal A_G=\int_{[0,1]^r}
 \prod_{e\in E(G)}
 2B_2\!\left(\{(C_G\bm p)_e\}\right)\dd^r p.
 \label{eq:graphintegral}
\end{equation}
The empty-dimensional integral for a tree is understood as evaluation
at zero. Neither the graph multiplicity nor the field cumulants are
included in $\mathcal A_G$. This establishes the exact connection
between the local-source expansion and the rational Bernoulli integrals,
without assuming distinct physical sites for distinct graph vertices.

\medskip
\noindent\textbf{Rationality of the field-expansion coefficients.}  Let
\begin{equation}
 y=\tanh^2h.
\end{equation}
Then, coefficient by coefficient in $x$,
\begin{equation}
 \phi(x,h)=\log(2\cosh h)+\sum_{n\ge1}\mathcal P_n(y)x^n,
 \qquad
 \mathcal P_n(y)\in\mathbb Q[y],
 \qquad
 \deg_y\mathcal P_n\le n.
 \label{eq:field-closure}
\end{equation}
Moreover,
\begin{equation}
 \mathcal P_1(y)=\frac{y}{6},
 \qquad
 \mathcal P_n(1)=0\quad(n\ge2),
 \label{eq:field-saturation}
\end{equation}
so that
\begin{equation}
 \mathcal P_n(y)=(1-y)\mathcal R_n(y),
 \qquad
 \mathcal R_n\in\mathbb Q[y],
 \qquad
 \deg\mathcal R_n\le n-1,
 \qquad n\ge2.
 \label{eq:field-factor}
\end{equation}

\medskip
\noindent The spatial integrals are rational for a simple reason:
each Bernoulli factor becomes a polynomial between successive integer
values of its argument. The following result states the needed fact
without assuming any special graph topology.

\begin{lemma}[Rational Bernoulli integrals]\label{lem:bern-rational}
Let $C$ be an integer matrix with $r$ columns, and let $F$ be a
polynomial with rational coefficients in its arguments. Then
\begin{equation}
 \int_{[0,1]^r}F\bigl(\{C\bm p\}\bigr)\,\dd^r p
 \in\mathbb Q,
 \label{eq:rational-grid-integral}
\end{equation}
where fractional parts are taken componentwise. This also includes
$r=0$, when the integral means evaluation of the integrand.
\end{lemma}
\begin{proof}
The conditions $(C\bm p)_e=j$, for integers $j$, divide the unit
cube into finitely many regions. On each region the integer parts are
fixed, so $F(\{C\bm p\})$ is a polynomial with rational coefficients.
Each region is bounded by hyperplanes with integer coefficients and
integer right-hand sides; its vertices therefore have rational
coordinates. Dividing such a region into simplices with rational
vertices shows that the integral of the polynomial is rational.
The boundaries have zero volume and do not affect the integral.
\end{proof}

\noindent\emph{Completion of the field-coefficient proof.}
Apply Lemma~\ref{lem:bern-rational} to the product of Bernoulli
factors in Eq.~\eqref{eq:graphintegral}. In particular, repeated edges,
multiple independent cycles, and intersections of the dividing
hyperplanes do not change rationality. This step needs only the integer
entries of $C_G$, not the stronger matrix property used for finite-ring
reconstruction.

By \eqref{eq:kappa-rec}--\eqref{eq:kappa-parity}, the local factor \eqref{eq:connected-partition-main} is an integer polynomial in $t$ of degree at most $\sum_vd_v=2n$ and parity $(-1)^{\sum_vd_v}=+1$.  It is therefore a polynomial in $y=t^2$ of degree at most $n$.  Summing the finitely many connected source graphs proves \eqref{eq:field-closure}.  At first order, the product measure has $\langle\sigma_i\sigma_j\rangle_h=t^2$, so
\begin{equation}
 \mathcal P_1(y)=\frac{y}{2}\sum_{r\ne0}\frac1{\pi^2r^2}=\frac{y}{6}.
\end{equation}
For $n\ge2$, every partition satisfying $\pi\vee\eta=\mathbf1$ has a block of size at least two; otherwise the $n$ bond groups remain disconnected. Since $\kappa_m(1)=0$ for $m\ge2$, every term of Eq.~\eqref{eq:connected-partition-main} vanishes separately at $t=1$. This proves \eqref{eq:field-factor} without requiring a cancellation among spatial graph amplitudes. The proof is coefficientwise and does not assume convergence of the entire high-temperature series.
\par

The zero-field statements are immediate corollaries.  Writing
\begin{equation}
 \phi(x,0)=\log2+\sum_{n\ge2}Q_nx^n,
 \qquad
 Q_n=\mathcal P_n(0),
 \label{eq:pressuredef}
\end{equation}
we have $Q_n\in\mathbb Q$.  More generally, define the connected zero-field magnetic cumulants by
\begin{equation}
 C_{2m}(x)=\left.\frac{\partial^{2m}\phi(x,h)}{\partial h^{2m}}\right|_{h=0},
\end{equation}
then
\begin{equation}
 [x^n]C_{2m}(x)
 =\left.\frac{\dd^{2m}}{\dd h^{2m}}\mathcal P_n(\tanh^2h)\right|_{h=0}
 \in\mathbb Q.
 \label{eq:field-sus-corollary}
\end{equation}
In particular, for the ordinary susceptibility,
\begin{equation}
 a_n^{(\chi)}=2\mathcal P_n'(0),
 \label{eq:A-from-field}
\end{equation}
where the prime denotes differentiation with respect to $y$. Thus at each perturbative order a single polynomial of degree at most $n$ determines the field dependence of $\phi=-\beta f$ and every derivative of the uniform magnetization at zero field.  In the original coupling $g=x/\pi^2$, every order-$n$ coefficient is a rational multiple of $\pi^{2n}$.

The theorem is an Ising-specific consequence of established Bernoulli mathematics.  Residue formulas, vector-partition Euler--Maclaurin theory, multiple Bernoulli series and root-system generalizations provide a natural language for the piecewise-polynomial periodic functions generated by the integral vector configuration $C_G$ \cite{Szenes1998,SzenesVergne2003,BoysalVergne2012,KomoriMatsumotoTsumura2010}.  The model-specific result is that each field-dependent high-temperature coefficient is a rational polynomial in $\tanh^2h$, with degree at most its perturbative order.

\subsection{Finite-ring polynomiality}
\label{sec:finiteL}

Rationality in infinite volume does not by itself provide an exact
formula for the finite-ring coefficients. For that we use the special
structure of the momenta carried by bonds in a connected graph. We first
show why those momenta have the required integer properties, and then
prove a finite-grid bound.

\medskip
\noindent\textbf{Momentum conservation on a graph.}
Orient the edges of a connected graph $G$ and remove one row from its
vertex--edge incidence matrix. Choose a spanning tree and write the
remaining matrix as $D=(D_T\mid D_C)$, where $D_T$ contains the tree
edges and $D_C$ the other edges. The square matrix $D_T$ has
determinant $\pm1$. Momentum conservation at each vertex gives
$D_T\bm k_T+D_C\bm k_C=0$, so the bond momenta may be written
in terms of the $r=|E|-|V|+1$ independent cycle momenta as
\begin{equation}
 C_G=\begin{pmatrix}-D_T^{-1}D_C\\ I_r\end{pmatrix},
 \qquad \bm k=C_G\bm k_C.
 \label{eq:network-cycle-matrix}
\end{equation}
The reduced incidence matrix $D$ has every square minor equal to
$0$ or $\pm1$. Indeed, in any square submatrix, a column with at most
one nonzero entry permits expansion by that column; if every column
has two entries, they have opposite signs and the rows sum to zero.
The same minor property holds for $-D_T^{-1}D_C$: a square minor
of this block equals, up to sign, the determinant obtained by replacing
the corresponding tree columns of $D_T$ by columns of $D_C$.
This is the standard network-matrix property. Adding the unit rows
$I_r$ preserves it. Thus every square minor of $C_G$ is $0$ or $\pm1$
(the property called \emph{total unimodularity}). Parallel bonds
cause no difficulty. Adding coordinate unit rows also preserves the
property, a fact needed to handle the boundaries of the momentum grid.

\medskip
\noindent Let $C\in\mathbb Z^{N\times r}$ have this property,
with row vectors $c_e$, and let $S\subseteq\{1,\ldots,N\}$ select
the Bernoulli factors. The $r$ columns represent independent momenta,
not magnetic fields. Define
\begin{equation}
 \mathcal T_{C,S}(L)=\frac1{L^r}\sum_{\bm k\in(\mathbb Z/L\mathbb Z)^r}
 \prod_{e\in S}2B_2\!\left(\left\{\frac{(C\bm k)_e}{L}\right\}\right).
 \label{eq:TCS}
\end{equation}
\begin{lemma}[Exact finite-grid bound]\label{lem:finite-grid}
Let $C\in\mathbb Z^{N\times r}$ have all square minors in
$\{0,\pm1\}$, and let $m=|S|$. At every integer $L\ge1$,
\begin{equation}
 \mathcal T_{C,S}(L)\in\mathbb Q[L^{-1}],\qquad
 \deg_{L^{-1}}\mathcal T_{C,S}(L)\le r+2m,
 \label{eq:TCS-weak}
\end{equation}
The constant term is the continuum integral of the same Bernoulli
product over the independent momenta $\bm p\in[0,1]^r$.
\end{lemma}
\begin{proof}
Divide $[0,1)^r$ where any $c_e\cdot\bm p$ passes through an
integer. The closures of the resulting regions are bounded by
integer-valued linear equations, including the coordinate boundaries.
A vertex is the solution of $r$ independent such equations. Because
all square minors of $C$ together with coordinate unit rows are
$0$ or $\pm1$, every vertex has integer coordinates. Thus these
regions have integral vertices.

On each region $Q$ the product of Bernoulli factors is an ordinary
polynomial with rational coefficients, of total degree at most $2m$.
Write it as a sum of homogeneous terms $f_d$, with $0\le d\le2m$.
The contribution from one term has the form
\begin{equation}
 \frac1{L^r}\sum_{\bm k\in LQ\cap\mathbb Z^r}
 f_d(\bm k/L)
 =L^{-r-d}\sum_{\bm k\in LQ\cap\mathbb Z^r}
 f_d(\bm k).
 \label{eq:weighted-grid-step}
\end{equation}
For a region with integral vertices, the weighted lattice-point
count on the right is a polynomial in $L$ of degree at most $r+d$
\cite{BeckRobins2015}. This is the polynomial-weight version of
ordinary lattice-point counting: no periodic dependence on $L$ occurs
because the vertices are integral. After the prefactor in
Eq.~\eqref{eq:weighted-grid-step}, the contribution is a polynomial
in $L^{-1}$ of degree at most $r+d$.

For completeness, grid points on a boundary must be counted only once.
Choose a disjoint convention for the shared faces, including the outer
faces at $p_j=1$ that are excluded from $[0,1)^r$. The sum over each
resulting partially open region equals its closed-region sum plus or
minus sums over selected faces and their intersections, by
inclusion--exclusion. These faces also have integral vertices; their
weighted lattice-point counts are polynomials in $L$ with degree no
larger than $r+d$. Moreover, $B_2(0)=B_2(1)$, so adjacent polynomial
expressions agree at shared boundaries. Thus the boundary convention
neither duplicates momentum points nor introduces periodic corrections.

Adding all regions and degrees $d\le2m$ proves
Eq.~\eqref{eq:TCS-weak}. Finally, as $L\to\infty$, the normalized
finite sum approaches the continuum momentum integral, since periodic
$B_2$ is continuous. This limit is the constant term of the finite
polynomial in $L^{-1}$.
\end{proof}

\noindent\emph{Independent conservative reconstruction.}
The preceding finite-grid lemma and the source-cumulant expansion give,
independently of any even-power cancellation,
\begin{equation}
 \mathcal P_{n,L}(y)=\sum_{j=0}^{3n}a_{n,j}(y)L^{-j},
 \qquad a_{n,j}(y)\in\mathbb Q[y],\quad \deg_y a_{n,j}\le n.
 \label{eq:ring-poly-weak}
\end{equation}
To extract the constant term, interpolate this polynomial in
$z=L^{-1}$ from $3n+1$ distinct values $z_a=L_a^{-1}$.
The Lagrange basis evaluated at $z=0$ is
$\prod_{b\ne a}(-z_b)/(z_a-z_b)
=L_a^{3n}/\prod_{b\ne a}(L_a-L_b)$, giving
\begin{equation}
 \mathcal P_n(y)=\sum_{a=0}^{3n}\mathcal P_{n,L_a}(y)
 \frac{L_a^{3n}}{\displaystyle\prod_{b\ne a}(L_a-L_b)}.
 \label{eq:field-interp-unconditional}
\end{equation}
This reconstruction follows solely from the finite-grid argument.
It also furnishes an independent modular check on the shorter
interpolation whenever sufficiently many ring sizes are available.

\subsection{Sharp even-power reconstruction}\label{sec:sharp-proof}
We now establish the sharp graphwise degree bound. The finite-grid lemma above supplies an independent polynomiality estimate, but it does not by itself give the required inverse-square degree. We use two complementary descriptions: rational functions on roots of unity for polynomiality and the Bernoulli Fourier kernel for parity. Throughout, the self-bond convention is $A_L(0)=0$.

\medskip
\noindent\textbf{One-root cyclotomic elimination.}
Write $\Omega_L=\{\zeta\in\mathbb C:\zeta^L=1\}$ and, for distinct roots, define
\begin{align}
 \mathscr K(a,b)&=\frac{a}{a-b},\qquad \mathscr K(a,b)+\mathscr K(b,a)=1,\nonumber\\
 L^2A_L(i-j)&=4\mathscr K(\zeta_i,\zeta_j)\mathscr K(\zeta_j,\zeta_i),
 \qquad \zeta_j=e^{2\pi i j/L}.
 \label{eq:root-kernel}
\end{align}
The displayed identity for $A_L$ is applied only to unequal physical sites; coincident bond endpoints give zero.

\begin{lemma}[Cyclotomic elimination]\label{lem:cyclotomic-elimination}
Fix distinct external roots $a_1,\ldots,a_m\in\Omega_L$. If $F(z)$ is a polynomial with rational coefficients in the generators $\mathscr K(z,a_i)$, $\mathscr K(a_i,z)$, and $\mathscr K(a_i,a_j)$ for $i\ne j$, then
\begin{equation}
 \sum_{z\in\Omega_L\setminus\{a_1,\ldots,a_m\}} F(z)
 \ \in\ \mathbb Q[L,\mathscr K(a_i,a_j):i\ne j],
 \label{eq:cyclotomic-closure}
\end{equation}
where the polynomial expression on the right depends only on $F$ and $m$, not on the selected roots or the value of $L$.
\end{lemma}
\begin{proof}
As a rational function of $z$, $F$ has poles only at the $a_i$ and is regular at infinity. Its partial fractions can therefore be written
\begin{equation}
 F(z)=F_\infty+\sum_{i=1}^{m}\sum_{q=1}^{q_i}
 C_{i,q}\left(\frac{z}{a_i}-1\right)^{-q}.
 \label{eq:cyclotomic-parts}
\end{equation}
Both $F_\infty$ and the coefficients $C_{i,q}$ are polynomials in the surviving generators $\mathscr K(a_i,a_j)$. To verify this, expand around $z=a_i(1+t)$: the two singular generators are $(1+t)/t$ and $-1/t$, while, for $j\ne i$, the regular ones have expansions
\begin{equation}
 \mathscr K(z,a_j)=\frac{\mathscr K(a_i,a_j)(1+t)}{1+\mathscr K(a_i,a_j)t},
 \qquad \mathscr K(a_j,z)=1-\mathscr K(z,a_j).
 \label{eq:cyclotomic-local}
\end{equation}
Every coefficient in the required Laurent expansion is a polynomial in the surviving generators.

For each fixed $q\ge1$, the cyclotomic sum $S_q(L)=\sum_{\omega\in\Omega_L\setminus\{1\}}(\omega-1)^{-q}$ is a polynomial over $\mathbb Q$ in $L$. Indeed,
\begin{equation}
 \sum_{\substack{\omega^L=1\\\omega\ne1}}
 \frac1{1+t-\omega}
 =\frac{L(1+t)^{L-1}}{(1+t)^L-1}-\frac1t,
 \qquad S_q(L)=-[t^{q-1}]\,H_L(t),
 \label{eq:cyclotomic-H}
\end{equation}
where $H_L(t)$ denotes the preceding rational function. For any fixed Taylor order, the binomial expansion of $(1+t)^L$ shows that the coefficient is polynomial in $L$ with rational coefficients. In the $i$th principal part of \eqref{eq:cyclotomic-parts}, summation over all $z\ne a_i$ gives $S_q(L)$. Excluding the other $a_j$ subtracts
$\sum_{j\ne i}(a_j/a_i-1)^{-q}=\sum_{j\ne i}[-\mathscr K(a_i,a_j)]^q$.
The constant term contributes $(L-m)F_\infty$. All contributions have the required form. The identity remains valid when the admissible sum is empty.
\end{proof}

\begin{lemma}[Polynomial graph amplitudes]\label{lem:cyclotomic-graphs}
For every connected loop-free multigraph $G$ with $n\ge1$ edges, including repeated edges, its unrestricted ring amplitude satisfies
\begin{equation}
 Q_G(L):=L^{2n}\mathcal A_{G,L}\in\mathbb Q[L]
 \qquad(L\ge1).
 \label{eq:graph-Q-polynomial}
\end{equation}
\end{lemma}
\begin{proof}
Use the position-space definition
\begin{equation}
 \mathcal A_{G,L}=\frac1L\sum_{s:V(G)\to\mathbb Z/L\mathbb Z}
 \prod_{uv\in E(G)}A_L(s_u-s_v).
 \label{eq:graph-spatial-appendix}
\end{equation}
Partition the vertex assignments according to which abstract graph vertices occupy the same physical site. If such a partition identifies the ends of any edge, its contribution vanishes. Every other partition gives a loop-free quotient multigraph on distinct occupied sites. By \eqref{eq:root-kernel}, its contribution to $Q_G(L)$ is a finite rational multiple of a sum of products of $\mathscr K$ generators over injectively assigned roots. Translation invariance fixes one root to $1$ and cancels the factor $L^{-1}$, so successive applications of Lemma~\ref{lem:cyclotomic-elimination} eliminate all the other roots and give a polynomial in $L$.

For completeness, the elimination formulas also apply at small rings. With $v$ distinct roots required, an injective sum is empty for $L<v$. Induction on the number of root variables proves that the polynomial obtained by elimination agrees with the empty sum: if $v-1>L$, the outer sum already vanishes; if $v-1=L$, the final restricted sum is empty and Lemma~\ref{lem:cyclotomic-elimination} gives zero. Thus no exceptional residue classes or small-$L$ corrections occur. Summing over the finitely many site partitions proves \eqref{eq:graph-Q-polynomial} for all $L$.
\end{proof}

\medskip
\noindent\textbf{Toric parity of the Fourier sums.}
Write $B_2(t)=t^2-t+1/6$, with $B_2(0)=B_2(1)$. The following reciprocity statement is the needed parity result; it does not presume a degree-$m$ bound for every individual Bernoulli product.
\begin{lemma}[Continuous toric grid reciprocity]\label{lem:toric-parity}
Let $C\in\mathbb Z^{N\times r}$ be totally unimodular, let $S\subseteq\{1,\ldots,N\}$, and put
$ f(\bm u)=\prod_{e\in S}2B_2(\{c_e\cdot\bm u\})$ on the torus $\mathbb R^r/\mathbb Z^r$. Then $\mathcal T_{C,S}(L)$ from \eqref{eq:TCS} is an even Laurent polynomial in $L$, with rational coefficients and no positive powers:
\begin{equation}
 \mathcal T_{C,S}(L)\in\mathbb Q[L^{-2}].
 \label{eq:TCS-parity}
\end{equation}
\end{lemma}
\begin{proof}
The hyperplanes $c_e\cdot\bm u\in\mathbb Z$, together with the coordinate hyperplanes $u_i\in\mathbb Z$, subdivide the torus into relatively open integral polyhedral cells. Their closures are integral: every vertex solves an integer linear system with coefficient rows from $C$ and the coordinate unit vectors, whose nonsingular square minors are $\pm1$. For a regular cell complex without self-identifications of closed cells, work on the finite covering torus $\mathbb R^r/(3\mathbb Z)^r$ and subdivide it by these hyperplanes; this changes the grid sum by the factor $3^r$ and leaves its normalized parity unchanged. The closures of its cells lie in translated unit cubes and have integral vertices.

On each relatively open $d$-cell $F$, the function $f$ is represented by a rational polynomial $p_F$ that extends to its closure and agrees there with $f$: this last assertion uses the continuity $B_2(0)=B_2(1)$ on every wall. Let $E_F^\circ(L)$ and $E_F(L)$ denote the weighted sums of $p_F(k/L)$ over lattice points in the dilated relative interior and closure of $F$. Weighted Ehrhart polynomiality and reciprocity for integral polytopes \cite{BeckRobins2015}, applied to each homogeneous part of $p_F$, give
\begin{equation}
 E_F^\circ(-L)=(-1)^{d}E_F(L).
 \label{eq:weighted-cell-reciprocity}
\end{equation}
Indeed, a degree-$h$ homogeneous weight contributes $L^{-h}$ times the usual degree-$h$ polynomial-weighted lattice sum; both the reciprocity sign $(-1)^{d+h}$ and the factor $(-L)^{-h}$ appear, leaving $(-1)^d$. The same argument applies to lower-dimensional faces, using their integral affine lattices.

Let $S^{(3)}(L)$ denote the unnormalized grid sum on the covering torus. Every point belongs to the relative interior of a unique cell, and hence
\begin{equation}
 S^{(3)}(-L)=\sum_F(-1)^{\dim F}E_F(L).
 \label{eq:toric-cell-sum}
\end{equation}
At a grid point in the relative interior of a cell $H$, all the polynomial extensions from cells $F\supseteq H$ have the same value $f$ by continuity. The coefficient of that value in \eqref{eq:toric-cell-sum} is $\sum_{F\supseteq H}(-1)^{\dim F}=(-1)^r$: the alternating coface sum is the local Euler/link identity for the regular decomposition of the closed $r$-torus. Therefore $S^{(3)}(-L)=(-1)^rS^{(3)}(L)$. Since $S^{(3)}(L)=3^r\sum_{\bm k\bmod L}f(\bm k/L)$, dividing by $L^r$ yields $\mathcal T_{C,S}(-L)=\mathcal T_{C,S}(L)$. Weighted Ehrhart polynomiality and the bound $S^{(3)}(L)=O(L^r)$ exclude positive powers of $L$, proving \eqref{eq:TCS-parity}. The case $r=0$ is immediate.
\end{proof}

\medskip
\noindent\textbf{Sharp connected-graph degree.}
\begin{lemma}[Sharp graph bound]\label{lem:graph-sharp}
Every connected loop-free multigraph $G$ with $n$ edges obeys
\begin{equation}
 \mathcal A_{G,L}=\sum_{j=0}^{n}q_{G,j}L^{-2j},
 \qquad q_{G,j}\in\mathbb Q.
 \label{eq:graph-sharp}
\end{equation}
\end{lemma}
\begin{proof}
The exact Fourier symbol is $\widehat A_L(k)=2B_2(k/L)-1/(3L^2)$. Expanding its product in \eqref{eq:finite-graph}, each term is a Bernoulli grid sum \eqref{eq:TCS} multiplied by an even nonpositive power of $L$. Lemma~\ref{lem:toric-parity} makes $\mathcal A_{G,L}$ an even Laurent polynomial. Hence, by Lemma~\ref{lem:cyclotomic-graphs}, $Q_G(L)=L^{2n}\mathcal A_{G,L}$ is an \emph{even ordinary polynomial} in $L$.

Let $v=|V(G)|$, and choose a spanning tree. The non-tree factors in $Q_G$ are bounded by $\csc^2(\pi r/L)\le L^2/4$ (for nonzero separations), and the tree factors can be summed successively from the leaves using $\sum_{r=1}^{L-1}\csc^2(\pi r/L)=(L^2-1)/3$. Every graph weight is nonnegative. Anchoring one vertex to cancel the factor $L^{-1}$ yields
\begin{equation}
 0\le Q_G(L)\le
 \left(\frac{L^2-1}{3}\right)^{v-1}
 \left(\frac{L^2}{4}\right)^{n-v+1}
 =O_G(L^{2n}).
 \label{eq:graph-growth}
\end{equation}
Thus $Q_G$ has ordinary degree at most $2n$. Its even polynomial form yields \eqref{eq:graph-sharp} after division by $L^{2n}$.
\end{proof}

\medskip
\noindent\textbf{Application to finite-ring field coefficients.}
The graph amplitudes also have the Fourier representation
\begin{equation}
 \mathcal A_{G,L}=\frac1{L^r}\sum_{\bm k\in(\mathbb Z/L\mathbb Z)^r}
 \prod_{e=1}^{n}\widehat A_L((C_G\bm k)_e).
 \label{eq:finite-graph}
\end{equation}
The finite-grid Lemma~\ref{lem:finite-grid} independently bounds the inverse-$L$ degree of each term selected from the Bernoulli and $-1/(3L^2)$ parts by $r+2n\le3n$. The graphwise Lemma~\ref{lem:graph-sharp} proves the stronger result for the full amplitude. The connected endpoint partitions in \eqref{eq:product-cumulant-graph} supply rational local cumulant polynomials in $y$, of degree at most $n$, independently of $L$. Thus the field coefficient is a finite rational polynomial combination of the amplitudes and satisfies
\begin{equation}
 \mathcal P_{n,L}(y)\in\mathbb Q[y,L^{-2}],\qquad
 \deg_{L^{-2}}\mathcal P_{n,L}\le n,
 \label{eq:finite-field-poly}
\end{equation}
including repeated bonds and coincident site assignments. The thermodynamic coefficient is its constant term because the corresponding graph limits exist by the Bernoulli Riemann sums.

Independently of finite-size reconstruction, deterministic polarization and the exact one-edge sum give
\begin{equation}
 \mathcal P_{1,L}(y)=\frac{y}{6}(1-L^{-2}),\qquad
 (1-y)\mid \mathcal P_{n,L}(y)\quad(n\ge2).
 \label{eq:finite-field-factor}
\end{equation}
In particular,
\begin{equation}
 \mathcal P_n(y)=\left.\mathcal P_{n,L}(y)\right|_{L^{-2}=0}.
 \label{eq:field-constant}
\end{equation}
For the susceptibility,
\begin{equation}
 \chi_L(x)=1+\sum_{n\ge1}a_{n,L}^{(\chi)}x^n,
 \qquad a_{n,L}^{(\chi)}=2\partial_y\mathcal P_{n,L}(0),
 \label{eq:AnLpoly}
\end{equation}
the same sharp polynomiality and exact $n+1$-ring reconstruction follow by differentiation. This completes the proof of Theorem~\ref{thm:ring}, except for the elementary Lagrange interpolation written explicitly below.

\subsection{Reconstruction and exact arithmetic}

The source recursion \eqref{eq:Wrecursion} is already an arbitrary-order coefficient generator. A second exact implementation, convenient for modular arithmetic, uses moments of the interaction energy and magnetization. On a ring set
\begin{equation}
 E_L(\sigma)=\frac12\sum_{i,j=0}^{L-1}A_{ij}\sigma_i\sigma_j,
 \qquad M(\sigma)=\sum_{i=0}^{L-1}\sigma_i,
 \label{eq:energy-M-definition}
\end{equation}
with the spin-independent self-image terms omitted. Write $\EE_0$ for averaging with the independent, unbiased spin measure (weight $2^{-L}$ per configuration). To distinguish this normalized expectation from the physical partition function $Z_L$, define
\begin{equation}
 \widetilde Z_L(x)=\EE_0 e^{xE_L}
 =\sum_{n\ge0}z_{n,L}x^n,
 \qquad
 N_L(x)=\frac1L\EE_0\!\left[M^2e^{xE_L}\right]
 =\sum_{n\ge0}m_{n,L}x^n,
\end{equation}
where
\begin{equation}
 z_{n,L}=\frac1{n!}\EE_0(E_L^n),
 \qquad
 m_{n,L}=\frac1{Ln!}\EE_0(M^2E_L^n).
 \label{eq:moment-coeffs}
\end{equation}
Since $\chi_L=N_L/\widetilde Z_L$ and $z_{0,L}=1$, the finite-ring susceptibility coefficients satisfy the triangular recursion
\begin{equation}
 a_{0,L}^{(\chi)}=1,\qquad
 a_{n,L}^{(\chi)}=m_{n,L}-\sum_{k=1}^{n}z_{k,L}a_{n-k,L}^{(\chi)}.
 \label{eq:finite-chi-rec}
\end{equation}
For the change in $\phi_L=-\beta f_L$ relative to $x=0$, define the finite-ring series
\begin{equation}
 \phi_L(x,0)-\log2=\frac1L\log \widetilde Z_L(x)=\sum_{n\ge1}Q_{n,L}x^n.
 \label{eq:finite-pressure-def}
\end{equation}
Differentiating $\log \widetilde Z_L$ or comparing coefficients gives
\begin{equation}
 Q_{n,L}=\frac{z_{n,L}}{L}
 -\frac1n\sum_{k=1}^{n-1}kQ_{k,L}z_{n-k,L},
 \qquad n\ge1.
 \label{eq:finite-pressure-rec}
\end{equation}
Equivalently, if $q_{n,L}=[x^n]\log \widetilde Z_L$, then $q_{n,L}$ obeys the same recurrence without the factors $1/L$ and $Q_{n,L}=q_{n,L}/L$.  This distinction is essential: the finite-size polynomiality theorem applies to the logarithm of the partition function per spin.  The physically irrelevant linear self-image contribution is removed according to the convention of Sec.~\ref{sec:model}.  Equations \eqref{eq:moment-coeffs}--\eqref{eq:finite-pressure-rec}, or equivalently the source recursion \eqref{eq:Wrecursion}, can be truncated at any prescribed order $N$.  There is no distinguished perturbative order at which the algebra changes.

For any chosen order $n$, Theorem~\ref{thm:ring} gives $a_{n,L}^{(\chi)}=F_n(L^{-2})$ with $\deg F_n\le n$. Lagrange interpolation at $u_a=L_a^{-2}$ gives the exact reconstruction formula:
\begin{equation}
 a_n^{(\chi)}=F_n(0)
 =\sum_{a=0}^{n}
 a_{n,L_a}^{(\chi)}\,
 \frac{L_a^{2n}}{\displaystyle\prod_{b\ne a}(L_a^2-L_b^2)} .
 \label{eq:closed-reconstruction}
\end{equation}
This formula is useful conceptually as well as computationally: it makes explicit that the thermodynamic limit is a finite algebraic functional of finite-volume data.  The uniform-field version uses the same interpolation weights:
\begin{equation}
 \mathcal P_n(y)=\sum_{a=0}^{n}
 \mathcal P_{n,L_a}(y)\,
 \frac{L_a^{2n}}{\displaystyle\prod_{b\ne a}(L_a^2-L_b^2)}.
 \label{eq:field-reconstruction}
\end{equation}
Thus finite ring calculations reconstruct the complete order-$n$ equation-of-state polynomial, not merely its zero-field derivatives.  The sharp theorem makes $n+1$ rings sufficient at every order; the independent $3n+1$-ring formula \eqref{eq:field-interp-unconditional} remains available as a consistency check.

For direct uniform-field reconstruction it is convenient to use the biased independent-spin functional
\begin{equation}
 \EE_t[f]=2^{-L}\sum_{\sigma}f(\sigma)\prod_{i=1}^{L}(1+t\sigma_i),
 \qquad t=\tanh h.
 \label{eq:biased-product}
\end{equation}
Then
\begin{equation}
 \widetilde Z_L(x,t)=\EE_t e^{xE_L}
\end{equation}
has factorial-normalized coefficients $z_{n,L}(t)=\EE_t(E_L^n)/n!$, and the same logarithmic triangular recurrence as \eqref{eq:finite-pressure-rec} returns $\mathcal P_{n,L}(t^2)$.  Because its degree in $y=t^2$ is at most $n$, $n+1$ distinct values of $y$ determine the full field polynomial at that order.  Together with $n+1$ ring sizes, a rectangular $(n+1)\times(n+1)$ grid of scalar finite-volume evaluations is therefore sufficient for a completely algebraic reconstruction.

A single spin enumeration per ring can do better by retaining magnetization sectors.  If
\begin{equation}
 H_{n,L}(M)=\sum_{\sigma:\,M(\sigma)=M}E_L(\sigma)^n,
\end{equation}
then
\begin{equation}
 z_{n,L}(t)=\frac1{2^L n!}\sum_M H_{n,L}(M)
 (1+t)^{(L+M)/2}(1-t)^{(L-M)/2}.
 \label{eq:mag-sector-field}
\end{equation}
Thus binning the energy moments by $M$ recovers the complete field polynomial from the same state enumeration used at zero field, rather than requiring separate runs at multiple $h$ values.  This gives a direct exact-arithmetic implementation of the uniform-field theorem.

Exact arithmetic is implemented prime by prime.  If $\omega$ is a primitive $L$th root of unity in a finite field,
\begin{equation}
 \frac1{L^2}\csc^2\frac{\pi r}{L}
 =-\frac{4\omega^r}{L^2(\omega^r-1)^2}.
 \label{eq:finitefieldkernel}
\end{equation}
Thus all moments in \eqref{eq:moment-coeffs} can be accumulated modulo primes $p$ for which $L\mid p-1$.  Different ring sizes need not share the same prime; using size-adapted primes avoids any growth of a global least common multiple.  Results from several primes are combined by the Chinese remainder theorem, and rational reconstruction is attempted only after the composite modulus is large enough.  A final prime not used in the reconstruction is reserved for validation.  This separation between the initial calculation and independent verification is important because the rational coefficients quickly become much larger than any individual machine-word prime.

An exact computational workflow at any prescribed order is: generate the truncated finite-ring series; obtain $a_{n,L}^{(\chi)}$ or the desired connected-source coefficient by triangular series algebra; evaluate it on any $n+1$ admissible ring sizes; apply \eqref{eq:closed-reconstruction}; combine enough prime-field images for rational reconstruction; and verify the result both at an unused ring and modulo unused primes. The direct spin-enumeration implementation scales exponentially with the largest ring and is not claimed to be asymptotically optimal. For substantially higher orders, the source recursion \eqref{eq:Wrecursion}, graph-isomorphism reduction, and memoized Bernoulli amplitudes provide a complementary computational strategy. The order-22 modular benchmark and the complete independently calculated field-dependent data through order 26 supply stringent checks on the finite-ring implementation and the interpolation procedure.

For the order-22 benchmark validation, the modulus was
\begin{equation}
 p=26771144401,
\end{equation}
with 67 as a primitive root.  Here $p-1$ is divisible by every $L=2,\ldots,26$.  At every order $n\le22$, interpolation from $L=2,\ldots,n+2$ predicts the unused $L=n+3$ value exactly modulo $p$.  The same validation prime also reproduces the rational thermodynamic coefficients after reduction modulo $p$.

\subsection{Independent computation of the field series through order 26}
The supplementary archive now supplies every rational coefficient of
$\mathcal P_n(y)$ through $n=26$.  The computation enumerates
rotational equivalence classes of spin configurations (binary necklaces) of length $L$, assigning each rotational class its orbit
multiplicity.  For a spin configuration encoded by a binary word, the
cyclic autocorrelation is $C_r=\sum_i\sigma_i\sigma_{i+r}$ and the
interaction energy is
\begin{equation}
 E_L=\frac12\sum_{r=1}^{L-1}A_L(r)C_r.
 \label{eq:magnetic-reproduction-energy}
\end{equation}
Energy moments are accumulated separately in every magnetization sector,
then converted to the biased generating function
\eqref{eq:mag-sector-field}; the connected coefficients follow by exact
power-series logarithms.  This avoids separate state enumerations for
different field values and retains every field power $y^m$ at once.

For each $1\le n\le26$, the thermodynamic field polynomial was reconstructed coefficientwise from $L=2,\ldots,n+2$ by the exact sharp interpolation formula \eqref{eq:field-reconstruction}. The theorem establishes the algebraic finite-size structure; the computational checks below separately test the finite-ring modular data and reconstructed rational entries.
Eighteen primes used for the initial calculation, all near $2^{61}$, were combined by the
Chinese remainder theorem and exact rational reconstruction.  Four
separately reserved primes validated each of the 377 reconstructed
rational coefficients; at each order they also validated every field
coefficient at the unused ring $L=n+3$, through $L=29$ at order 26. The archived ring data permit more extensive checks: for the 18 original primes we compared with every additional ring up to $L=28$, and for the four reserved primes with every additional ring up to $L=29$. All 73,008 coefficient-by-ring-by-prime comparisons agree with the short-interpolation predictions.

For a genuinely different finite-volume implementation, we also
enumerated all $2^L$ spin configurations for $L=3,5,7,9,11,12$.
This check computed energies directly from the inverse discrete Fourier
transform of the quadratic Bernoulli symbol, evaluated the partition
function at distinct field values, and interpolated in $y$ rather than
using necklace orbit weights or the magnetization-sector convolution.
All of its order-26 finite-ring coefficients agreed with the primary implementation. The distinction is that this second implementation independently tests the finite-volume calculation, whereas the four reserved primes test the modular reconstruction without reusing the original primes. The resulting reconstructed bulk rows also agree exactly with the separately
archived susceptibility coefficients $a_n^{(\chi)}=2\mathcal P_n'(0)$ (denoted $A_n$ in the unchanged supplementary code and data), the
highest-degree and first-subleading coefficient formulas, and the coefficient identities
reported for the second and further subleading sequences.  The higher-collision-index
formulae remain conjectural beyond the checked orders.

The independent $3n+1$-ring interpolation supplied by Lemma~\ref{lem:finite-grid} provides a further check without using the sharp degree bound. It requires $3n+1$ rings and can be evaluated from the archived data through $n=8$ with all 18 calculation primes, and through $n=9$ with the four reserved primes. For each available prime, its results agree coefficient by coefficient with the reported rational values reduced modulo that prime (44 coefficients through $n=8$ and 10 more at $n=9$). These congruences are independent of the even-power interpolation rule, but do not by themselves certify the rational numbers over $\mathbb Q$ without suitable size bounds. Source code, individual modular ring outputs, exact coefficient files and runnable verification tests are included in the supplementary package.

\section{Proof of the highest-degree coefficient formula from the matching determinant}\label{app:A-matching}
\label{sec:topfield}

For later use, express the finite-ring quantity $\phi_L=-\beta f_L$ in terms of the same field variable:
\begin{align}
 \phi_L(x,h)&=\log2+\frac1L W_L(x,h\one)\nonumber\\
 &=\log(2\cosh h)+\sum_{n\ge1}\mathcal P_{n,L}(y)x^n,
 \qquad y=\tanh^2h.
 \label{eq:finite-field-pressure}
\end{align}
The degree bound in \eqref{eq:field-closure} leaves one distinguished coefficient at each order, namely
\begin{equation}
 d_{0,L}(n)=[y^n]\mathcal P_{n,L}(y),
 \qquad
 d_0(n)=[y^n]\mathcal P_n(y).
 \label{eq:topfield-def}
\end{equation}
This highest-degree field-polynomial contribution is exactly summable. The mechanism is a weighted monomer--dimer reduction \cite{HeilmannLieb1972} followed by a Cauchy-type determinant identity; related Cauchy-matrix methods are reviewed in \cite{Forrester2010}.

For the finite ring define the skew matrix
\begin{equation}
 (K_L)_{ij}=\begin{cases}
 \displaystyle \frac1L\csc\frac{\pi(i-j)}L,&i\ne j,\\[2mm]
 0,&i=j,
 \end{cases}
 \qquad i,j=0,\ldots,L-1,
 \label{eq:KL-csc}
\end{equation}
so that $(K_L)_{ij}^2=A_L(i-j)$ off the diagonal.  Let
\begin{equation}
 \mathfrak M_L(w)=
 \sum_{\mathcal M}w^{|\mathcal M|}
 \prod_{\{i,j\}\in\mathcal M}A_L(i-j),
 \label{eq:matching-partition}
\end{equation}
where the sum runs over all matchings of the complete graph on the $L$ ring sites, including the empty matching.

\medskip
\noindent\textbf{Matching determinant.}  For every $L\ge2$,
\begin{equation}
 \mathfrak M_L(w)=\det(I+\sqrt w\,K_L)
 =\prod_{k=0}^{\lfloor L/2\rfloor-1}
 \left[1+w\left(\frac{L-1-2k}{L}\right)^2\right].
 \label{eq:matching-determinant}
\end{equation}
Consequently,
\begin{equation}
 d_{0,L}(n)=\frac{(-1)^{n+1}}{2nL}
 \sum_{k=0}^{L-1}
 \left(\frac{L-1-2k}{L}\right)^{2n},
 \label{eq:topfield-finite}
\end{equation}
and the thermodynamic coefficient is
\begin{equation}
 d_0(n)=[y^n]\mathcal P_n(y)
 =\frac{(-1)^{n+1}}{2n(2n+1)}.
 \label{eq:topfield-coefficient}
\end{equation}
Equivalently, the generating function for the highest-degree coefficients is
\begin{align}
 \mathcal C(w)
 &:=\sum_{n\ge1}d_0(n)w^n
 =\frac14\int_{-1}^{1}\log(1+wu^2)\dd u\nonumber\\
 &=\frac{\arctan\sqrt w}{\sqrt w}-1+\frac12\log(1+w),
 \label{eq:topfield-generating}
\end{align}
understood as a formal series at $w=0$.

\par\noindent\emph{Derivation.}
Start from the finite-ring interaction partition function in the biased product measure,
\begin{equation}
 \widetilde Z_L(x,t)=
 2^{-L}\sum_{\sigma}e^{xE_L(\sigma)}
 \prod_{i=0}^{L-1}(1+t\sigma_i).
 \label{eq:field-Z-diagonal}
\end{equation}
At order $x^n$ its degree in $t$ is at most $2n$.  Introduce the diagonal extractor
\begin{equation}
 \mathcal D F(w)=\sum_{n\ge0}[x^nt^{2n}]F(x,t)\,w^n.
 \label{eq:diagonal-extractor}
\end{equation}
On series obeying this degree bound, $\mathcal D(FG)=\mathcal D(F)\mathcal D(G)$, since a product can attain the maximum field degree $2n$ at order $x^n$ only if every factor does so at its own order. Hence $\mathcal D\log F=\log\mathcal D F$ for such series with $F(0,t)=1$, with the logarithm understood as a formal power series.

In \eqref{eq:field-Z-diagonal}, the coefficient of $x^nt^{2n}$ comes from $n$ interaction bonds with $2n$ distinct endpoints, hence a matching. The $n!$ orderings of these bonds cancel the factor $1/n!$ from the exponential. This statement concerns the partition function before taking its logarithm. The logarithm then accounts for connected subtractions, including terms involving coincident sites in a direct cumulant expansion. Thus
\begin{equation}
 \mathcal D\widetilde Z_L=\mathfrak M_L,
 \qquad
 \sum_{n\ge1}d_{0,L}(n)w^n=\frac1L\log\mathfrak M_L(w).
 \label{eq:diagonal-to-matching}
\end{equation}

It remains to evaluate the matching partition function.  We use the following elementary Cauchy determinant--Hafnian lemma.  Let $C$ be an even-dimensional skew matrix of the form
\begin{equation}
 C_{ab}=\frac{u_au_b}{\omega_a-\omega_b},\qquad C_{aa}=0.
\end{equation}
Then
\begin{equation}
 \det C=\operatorname{Hf}(C^{\circ2}),
 \label{eq:cauchy-det-haf}
\end{equation}
where $C^{\circ2}$ denotes the entrywise square. For a symmetric matrix $M$ of size $2q$, its Hafnian is the sum over perfect pairings of the indices,
\begin{equation}
 \operatorname{Hf}(M)=\sum_{\text{perfect pairings }\mathcal M}
 \prod_{\{a,b\}\in\mathcal M}M_{ab}.
 \label{eq:hafnian-defined}
\end{equation}
Thus the Hafnian is simply the total weight of all perfect matchings. To prove \eqref{eq:cauchy-det-haf}, expand the determinant in permutation cycles. Two-cycles give the desired pairings. For any fixed set $S$ of $m\ge3$ vertices, the longer directed cycles contribute, after factoring out $\prod_{a\in S}u_a^2$, the symmetric rational function
\begin{equation}
 \sum_{\text{directed cycles on }S}
 \prod_{(a,b)}\frac1{\omega_a-\omega_b}.
\end{equation}
Multiplying by the Vandermonde $\Delta_S=\prod_{a<b}(\omega_a-\omega_b)$ gives an antisymmetric polynomial of degree $\binom m2-m$, strictly less than $\deg\Delta_S=\binom m2$. Every nonzero antisymmetric polynomial is divisible by $\Delta_S$, so this polynomial must vanish. Thus all longer cycles cancel, leaving only two-cycles. For a transposition $(ab)$, the negative permutation sign cancels the minus sign in $C_{ab}C_{ba}=-C_{ab}^2$. The remaining products are exactly the perfect pairings defining the Hafnian, proving \eqref{eq:cauchy-det-haf}.  This identity is in the same Cauchy/Pfaffian--Hafnian family as the classical identities reviewed in Ref.~\cite{IshikawaEtAl2005}.

Set $\omega_j=e^{2\pi\mathrm i j/L}$ and choose the square-root phases consistently as $\omega_j^{1/2}=e^{\pi\mathrm i j/L}$, where $\mathrm i^2=-1$. These are not independently chosen principal square roots. The cosecant matrix \eqref{eq:KL-csc} then has the Cauchy form
\begin{equation}
 (K_L)_{ij}=\frac{2\mathrm i}{L}
 \frac{\omega_i^{1/2}\omega_j^{1/2}}{\omega_i-\omega_j}.
\end{equation}
Hence every even principal minor satisfies
\begin{equation}
 \det(K_L)_S=\operatorname{Hf}\!\left((A_L)_S\right).
\end{equation}
The principal-minor expansion now gives
\begin{equation}
 \det(I+sK_L)
 =\sum_{S}s^{|S|}\det(K_L)_S
 =\sum_{\mathcal M}s^{2|\mathcal M|}
 \prod_{e\in\mathcal M}A_e,
\end{equation}
which is the first equality in \eqref{eq:matching-determinant} after setting $w=s^2$.

The physical interaction $A_L$ is periodic, but the signed auxiliary kernel of $K_L$ changes sign when its separation is increased by $L$, since $\csc(\theta+\pi)=-\csc\theta$. It is therefore diagonalized by antiperiodic Fourier modes, satisfying $\psi_{j+L}^{(k)}=-\psi_j^{(k)}$:
\begin{equation}
 \psi_j^{(k)}=e^{\pi\mathrm i(2k+1)j/L},
 \qquad k=0,\ldots,L-1.
\end{equation}  The required finite sum is
\begin{equation}
 \sum_{r=1}^{L-1}
 \frac{e^{\pi\mathrm i(2k+1)r/L}}{\sin(\pi r/L)}
 =\mathrm i(L-1-2k),
 \label{eq:csc-half-sum}
\end{equation}
where the real part cancels under $r\leftrightarrow L-r$, while the imaginary part follows from
$\sin((2k+1)x)/\sin x=1+2\sum_{m=1}^{k}\cos(2mx)$.  Thus
\begin{equation}
 \lambda_k(K_L)=-\mathrm i\frac{L-1-2k}{L}.
 \label{eq:KL-spectrum}
\end{equation}
Pairing opposite eigenvalues gives the product in \eqref{eq:matching-determinant}.  Expanding its logarithm yields \eqref{eq:topfield-finite}; the midpoint Riemann sum tends to
\begin{equation}
 \frac1L\sum_{k=0}^{L-1}
 \left(\frac{L-1-2k}{L}\right)^{2n}
 \longrightarrow \frac1{2n+1},
\end{equation}
which proves \eqref{eq:topfield-coefficient} and \eqref{eq:topfield-generating}.
\par

This closed-form result goes beyond the general rationality statement: it gives one coefficient of every field polynomial in closed form and supplies an independent all-order check on both the field-degree bound and the finite-size reconstruction.

\section{Exact source determinant at fixed vertex-collision index}\label{app:A-defect}
The matching determinant can be extended to configurations containing vertices at which bond endpoints coincide.  The source construction below packages these insertions at any fixed vertex-collision index, before the first two nontrivial cases are evaluated in Appendices~\ref{app:A-d1} and \ref{app:A-d2}. It is an exact finite-ring reduction; the conjectured rational formulas for further subleading coefficients are separate statements. Here $w$ marks interaction edges, while the auxiliary variable $u$ marks vertex-collision index; neither is a magnetic field.

The diagonal substitution $x=wu$, $t=u^{-1/2}$ assigns to a multigraph with $n$ edges and $o(G)$ odd-degree vertices the power $u^{n-o(G)/2}$.  Since $2n=\sum_vd_v$,
\begin{equation}
 n-\frac{o(G)}2=\sum_v\left\lfloor\frac{d_v}{2}\right\rfloor,
\end{equation}
which establishes the vertex-collision index in Eq.~\eqref{eq:defect-grade-main}.  Introduce a formal Wick functional on symbols $X_i$, defined by pair contractions $\langle X_iX_j\rangle=wA_{ij}$ and Wick's pairing rule, and
\begin{equation}
 f_u(X)=\sum_{d\ge0}u^{\lfloor d/2\rfloor}\frac{X^d}{d!}
 =\cosh(\sqrt u X)+\frac1{\sqrt u}\sinh(\sqrt u X).
\end{equation}
Then the complete ring partition function organized by vertex-collision index is
\begin{equation}
 \mathcal F_L(w,u)=\left\langle\prod_{i=1}^Lf_u(X_i)\right\rangle.
\end{equation}
To see the correspondence with the Ising expansion, expand each $f_u(X_i)$ and pair all $X$ factors. Every pair is an interaction bond with weight $wA_{ij}$, while a site incident on $d_i$ bonds carries $u^{\lfloor d_i/2\rfloor}$. The resulting weight is $w^n u^{k(G)}$, exactly the one obtained from the normalized spin partition function by the diagonal substitution above. The Wick notation is therefore a formal way of collecting the original bond configurations, not an additional approximation or a positive-covariance assumption.
At $u=0$, $f_0(X)=1+X$ and the matching determinant is recovered. More generally, for arbitrary site weights $a_i,b_i$, write $D_a=\operatorname{diag}(a_1,\ldots,a_L)$ and $D_b=\operatorname{diag}(b_1,\ldots,b_L)$. Then
\begin{equation}
 \left\langle\prod_i(a_i+b_iX_i)\right\rangle
 =\det(D_a+\sqrt w K_LD_b).
 \label{eq:abdet}
\end{equation}
Introduce independent auxiliary sources $\bm\eta=(\eta_1,\ldots,\eta_L)$ and site weights $\bm b=(b_1,\ldots,b_L)$. They are distinct from the physical reduced field $h$. Writing $\Gamma=wA$ for the Wick covariance, and $D_b=\operatorname{diag}(b_i)$, Gaussian shifting gives
\begin{equation}
 F_L(\bm\eta,\bm b)=
 \left\langle e^{\bm\eta\cdot X}\prod_i(1+b_iX_i)\right\rangle
 =e^{\frac12\bm\eta^T\Gamma\bm\eta}
 \det[D(\bm\eta,\bm b)+\sqrt w K_LD_b],
 \label{eq:sourcedet}
\end{equation}
where $D(\bm\eta,\bm b)_{ii}=1+b_i(\Gamma\bm\eta)_i$.  If $S$ is the exceptional set, choose $b_i=0$ on $S$ and $b_i=1$ elsewhere. Then
\begin{equation}
 \left.\prod_{i\in S}\partial_{\eta_i}^{d_i}
 F_L(\bm\eta,\bm b)\right|_{\bm\eta=0}
 =\left\langle\prod_{i\in S}X_i^{d_i}
 \prod_{j\notin S}(1+X_j)\right\rangle.
\end{equation}
An insertion of local vertex-collision index $\ell_i$ is therefore produced by the bounded local operator
\begin{equation}
 B_{\ell_i}(\partial_{\eta_i})=
 \frac{\partial_{\eta_i}^{2\ell_i}}{(2\ell_i)!}
 +\frac{\partial_{\eta_i}^{2\ell_i+1}}{(2\ell_i+1)!}.
\end{equation}
At fixed vertex-collision index $k$, at most $k$ sites are exceptional, with $\sum_i\ell_i=k$. Applying their local operators to Eq.~\eqref{eq:sourcedet} and then taking the logarithm gives an exact finite-ring procedure. This avoids enumerating all interaction multigraphs separately, but still requires evaluating source derivatives and connected subtractions. It has not been shown that these operations yield the conjectured rational coefficients for general $k$.

\section{Proof of the first subleading coefficient formula by cycle-Laplacian reduction}\label{app:A-d1}
The matching determinant settles the coefficient of $y^n$.  For $y^{n-1}$ one exceptional vertex remains, and the task is to sum its allowed local stars over a matching background.  The argument below converts the deleted-matching sums into a determinant quotient and then into a finite cycle-Laplacian problem.  This is the analytic input behind Theorem~\ref{thm:d1}.

We divide the proof into three operations: enumerate the allowed star insertions over a matching background, rewrite their deleted matching minors as a determinant quotient, and diagonalize that quotient using a cycle Laplacian. We keep every step at finite $L$ until the final resolvent limit.

Let $\mathfrak M_L(w)=\det(I+\sqrt w K_L)$ be the matching determinant and put $s=\sqrt w$, $B=I+sK_L$.  Write $B_0=B[\{1,\ldots,L-1\},\{1,\ldots,L-1\}]$ for the principal submatrix obtained by deleting the row and column of site $0$.  Fix the exceptional vertex at site $0$ and define $W=\operatorname{diag}(A_{0j})_{j=1}^{L-1}$ with $A_{0j}=K_{0j}^2$.  A configuration with vertex-collision index one has precisely one exceptional vertex, of degree two or three, with all other nonisolated vertices forming a matching background.  Fixing this exceptional vertex removes the overall factor of $L$ from translations.  For each choice of its distinct leaves, the remaining matching partition function is a principal deleted minor of $B$; Jacobi's complementary-minor theorem converts its ratio to $\det B$ into a complementary minor of $B^{-1}$.  For a subset $T\subseteq\{1,\ldots,L-1\}$ of $r$ distinct leaves, the principal-minor expansion and Jacobi's identity give
\begin{equation}
 \begin{aligned}
 [\lambda^r]\det(B_0+\lambda W)
 &=\sum_{|T|=r}\biggl(\prod_{j\in T}A_{0j}\biggr)
 \det B[(\{0\}\cup T)^c],\\
 \frac{\det B[U^c]}{\det B}&=\det(B^{-1})[U].
 \end{aligned}
 \label{eq:star-deleted-minors}
\end{equation}
Here $B[U^c]$ means the principal submatrix retained after deleting the sites in $U$, and $B^{-1}[U]$ is the principal submatrix on $U$. This makes the matching-background sum explicit. Introduce an auxiliary variable $\lambda$ (not the field variable $t=\tanh h$) and define
\begin{equation}
 \Phi_L(\lambda;w)=\frac{\det(B_0+\lambda W)}{\mathfrak M_L(w)},
 \qquad D_L^{(r)}(w):=L \mathfrak M_L(w)w^r[\lambda^r]\Phi_L,
 \quad r=2,3.
 \label{eq:star-sector-definitions}
\end{equation}
Thus $D_L^{(2)}$ and $D_L^{(3)}$ are respectively the two- and three-leaf star sums with their powers of $w$ and the $L$ choices of exceptional site restored. They are the only contributions at vertex-collision index one.

Set $\theta_j=\pi j/L$, $S=\operatorname{diag}(\sin\theta_j)$, and $\xi_j=\cot\theta_j$. Put $H_L:=SB_0S$, an $(L-1)\times(L-1)$ matrix. Because $A_{0j}=1/(L^2\sin^2\theta_j)$,
\begin{equation}
 S(B_0+\lambda W)S=H_L+\frac{\lambda}{L^2}I.
\end{equation}
To specify the interpolation map, set
\begin{equation}
 P_L(X)=\operatorname{Im}(X+i)^L,
 \qquad P_L(\xi_j)=0,
 \qquad P_L'(\xi_j)=L(-1)^{j+1}\sin^{2-L}\theta_j .
 \label{eq:interpolation-roots}
\end{equation}
For a polynomial $f$ of degree at most $L-2$, put
\begin{equation}
 g_f(\theta)=\sin^{L-1}\theta\,f(\cot\theta),
 \qquad U_{jj}=\frac{1}{P_L'(\xi_j)\sin^{L-1}\theta_j}
             =\frac{(-1)^{j+1}}{L\sin\theta_j} .
 \label{eq:interpolation-gauge}
\end{equation}
The nodal vector representing $f$ in the matrix $H_L$ is
\emph{$Ug_f$, not $g_f$}.  This diagonal gauge is essential for the sign and
normalization of the operator formula.  Indeed, the elementary identity
\begin{equation}
 \frac{\sin\theta_j\sin\theta_k}{\sin(\theta_j-\theta_k)}
      =\frac{1}{\xi_k-\xi_j}
 \qquad(j\ne k)
\end{equation}
and the Lagrange differentiation matrix
\begin{equation}
 D_{jk}=\frac{P_L'(\xi_j)}{P_L'(\xi_k)(\xi_j-\xi_k)}\quad(j\ne k),
 \qquad D_{jj}=\frac{P_L''(\xi_j)}{2P_L'(\xi_j)}
               =(L-1)\sin\theta_j\cos\theta_j
\end{equation}
show, by substituting $f'=\sum_kD_{jk}f(\xi_k)$, that
\begin{equation}
 H_LU\,\bm g_f=U\,\bm{T_Lg_f},
 \qquad T_L=\sin^2\theta\left(1+\frac{s}{L}\frac{\dd}{\dd\theta}\right).
 \label{eq:HL-intertwining}
\end{equation}
Here boldface denotes evaluation at $\theta_1,\ldots,\theta_{L-1}$.
The derivation is an equality on every polynomial $f$ in the indicated
$(L-1)$-dimensional space, and therefore a genuine similarity relation,
not merely a comparison of characteristic polynomials.

The functions $g_f$ are linear combinations of the parity Fourier modes
$e^{iq_m\theta}$, $q_m=L-1-2m$ ($m=0,\ldots,L-1$), whose coefficients sum
to zero since $g_f(0)=0$.  This identifies the interpolation space with the
sum-zero subspace of $\mathbb C^L$. Define the cycle Laplacian, with indices understood modulo $L$, by
\begin{equation}
 (\Lambda_L)_{m\ell}=2\delta_{m\ell}
 -\delta_{m,\ell+1\ (\mathrm{mod}\ L)}
 -\delta_{m,\ell-1\ (\mathrm{mod}\ L)}.
 \label{eq:cycle-laplacian-definition}
\end{equation}
Multiplication by $\sin^2\theta$ acts on the parity Fourier modes as one quarter of $\Lambda_L$, while the derivative factor is diagonal. In this Fourier representation,
\begin{equation}
 T_L=\left.\frac14\Lambda_LG_L\right|_{\sum v_m=0},
 \qquad G_L=\operatorname{diag}(g_m),
 \qquad g_m=1+i\frac{s}{L}q_m.
 \label{eq:cycle-fourier-operator}
\end{equation}
The sum-zero space is invariant because every column of $\Lambda_L$ sums to zero.  Since $\Lambda_LG_L$ has one zero eigenvalue, its full
determinant obeys
\begin{equation}
 \det(H_L+aI_{L-1})
    =\frac1a\det\left(aI_L+\frac14\Lambda_LG_L\right).
 \label{eq:cycle-characteristic}
\end{equation}
The standard sine product $\prod_{j=1}^{L-1}\sin^2(\pi j/L)
=L^2/4^{L-1}$ and $\prod_mg_m=\mathfrak M_L(w)$ then imply the exact finite-ring
identity
\begin{equation}
 [\lambda^r]\Phi_L(\lambda;w)=\frac{4^r}{L^{2r+2}}
  \sum_{\substack{T\subset\{0,\ldots,L-1\}\\ |T|=r+1}}
   \frac{\det\Lambda_L[T^c]}{\prod_{m\in T}g_m},
 \qquad 0\le r\le L-1.
 \label{eq:defect-cycle-minors}
\end{equation}
It follows from the principal-minor expansion of
Eq.~\eqref{eq:cycle-characteristic} and is independent of any continuum limit.
For $r+1$ deleted cycle vertices $m_1<\cdots<m_{r+1}$, the
matrix-tree theorem gives
\begin{equation}
 \det\Lambda_L[T^c]=\prod_{a=1}^{r+1}(m_{a+1}-m_a),
 \qquad m_{r+2}=m_1+L.
 \label{eq:cycle-gap-minor}
\end{equation}
Consequently every leaf sector is an explicit finite sum over cyclic gaps. For example, the two-leaf sector is
\begin{equation}
 \frac{D_L^{(2)}}{L\mathfrak M_L(w)}
 =\frac{16w^2}{L^6}
 \sum_{0\le a<b<c<L}
 \frac{(b-a)(c-b)(L-c+a)}{g_ag_bg_c}.
 \label{eq:two-leaf-gap-sum}
\end{equation}
The three-leaf sector similarly involves four deleted cycle vertices and the product of their four cyclic gaps, with prefactor $64w^3/L^8$. Both finite sums can be reduced without taking a large-$L$ limit. Set
\begin{equation}
 q_m=L-1-2m,\qquad
 \delta=g_{m+1}-g_m=-\frac{2is}{L}.
\end{equation}
For any arithmetic progression $g_m$ with common difference $\delta$, partial fractions in the denominators and elementary sums over the remaining gap positions give
\begin{align}
 &\sum_{a<b<c}\frac{(b-a)(c-b)(L-c+a)}{g_ag_bg_c}
 =\frac{L}{24\delta^2}\sum_m\frac{3q_m^2-L^2+1}{g_m},\nonumber\\
 &\sum_{a<b<c<d}\frac{(b-a)(c-b)(d-c)(L-d+a)}{g_ag_bg_cg_d}
 =\frac{L}{288\delta^3}
 \sum_m\frac{q_m(5q_m^2-3L^2+7)}{g_m}.
 \label{eq:gap-single-sums}
\end{align}
For clarity, the first identity follows by decomposing the summand as
\begin{equation}
\begin{aligned}
&\frac{(b-a)(c-b)(L-c+a)}{g_ag_bg_c}\\
&=\frac{L-c+a}{\delta^2}
\left[\frac{c-b}{c-a}\frac1{g_a}-\frac1{g_b}
+\frac{b-a}{c-a}\frac1{g_c}\right].
\end{aligned}
\end{equation}
Summing the coefficients of each $1/g_m$ gives the first line of Eq.~\eqref{eq:gap-single-sums}; the four-index partial fractions give the second line in the same way. These expressions show explicitly how the many-site sums reduce to one momentum sum. Define
\begin{equation}
 R_L(w)=\frac1L\sum_m\frac1{g_m}
       =\frac1L\operatorname{Tr}(I+sK_L)^{-1}.
\end{equation}
Since $g_m=1+isq_m/L$ and $\sum_mq_m=0$, polynomial division gives, for example,
\begin{equation}
\sum_m\frac{q_m^2}{g_m}=-\frac{L^3}{w}(R_L-1),
\qquad \sum_mq_m^2=\frac{L(L^2-1)}3.
\end{equation}
The corresponding identity for $q_m^3/g_m$ follows by the same division. Substituting into Eq.~\eqref{eq:gap-single-sums} gives the two finite-ring star contributions:
\begin{align}
 \frac{D_L^{(2)}}{L\mathfrak M_L(w)}&=\frac12(R_L-1)+\frac{L^2-1}{6L^2}wR_L,\\
 \frac{D_L^{(3)}}{L\mathfrak M_L(w)}&=-\frac5{36}(R_L-1)+\frac{-3+7L^{-2}}{36}wR_L
 +\frac{1-4L^{-2}}{27}w.
\end{align}
Adding the sectors gives
\begin{equation}
 \frac{D_L^{(2)}+D_L^{(3)}}{L\mathfrak M_L(w)}
 =\frac{13}{36}(R_L-1)+\frac{3+L^{-2}}{36}wR_L
 +\frac{1-4L^{-2}}{27}w.
 \label{eq:d1-finite-ring}
\end{equation}
The exact half-integer spectrum implies
\begin{equation}
 R_L(w)\longrightarrow \frac12\int_{-1}^1\frac{\dd u}{1+wu^2}
 =\frac{\arctan\sqrt w}{\sqrt w}.
\end{equation}
Therefore
\begin{equation}
 G_1(w)=\sum_{n\ge2}d_1(n)w^n
 =\frac1{36}\left[(13+3w)\frac{\arctan\sqrt w}{\sqrt w}-13+\frac43w\right].
\end{equation}
Indeed, the coefficient of $w^n$ for $n\ge2$ is
\begin{equation}
 [w^n]G_1(w)=\frac{(-1)^n}{36}
 \left(\frac{13}{2n+1}-\frac3{2n-1}\right)
 =(-1)^n\frac{5n-4}{9(2n-1)(2n+1)},
\end{equation}
which proves Eq.~\eqref{eq:d1-main}.  The derivation is finite-ring until the last elementary Riemann-sum limit.

\clearpage
\section{Second subleading coefficient and a non-series-parallel graph check}\label{app:A-d2}
The second subleading coefficient illustrates how the source determinant constrains the possible configurations with shared bond endpoints, while the simplification of the resulting connected sum remains conjectural.  Here $\mathcal F_L(w,u)$ denotes the Wick partition function organized by vertex-collision index, defined in Appendix~\ref{app:A-defect}: its variable $u$ counts vertex-collision index, so no further substitution of the coupling and field is made.  Put
\begin{align}
 G_{2,L}(w)&=[u^2]\,L^{-1}\log \mathcal F_L(w,u),\nonumber\\
 G_2(w)&=\lim_{L\to\infty}G_{2,L}(w)=\sum_{n\ge2}d_2(n)w^n .
\end{align}
At vertex-collision index two, the possible configurations are finite in number.  There is either one exceptional vertex of degree four or five, or two exceptional vertices with degree pairs
\begin{equation}
 (2,2),\quad(2,3),\quad(3,2),\quad(3,3),
\end{equation}
and the remaining vertices are degree-one matching vertices.  For
\begin{equation}
 \mathsf R_L=(I+\sqrt w K_L)^{-1},
 \qquad
 \frac{\mathfrak M_{L\setminus S}(w)}{\mathfrak M_L(w)}=\det\mathsf R_L[S],
\end{equation}
Jacobi's complementary-minor identity reduces these contributions to deleted matching minors involving at most eight sites. The maximum occurs for two degree-three exceptional sites, each attached to three distinct degree-one sites; these eight sites must be excluded from the remaining matching background. This bound describes the insertions before the connected subtraction, not the number of matrix operations needed after simplification. To specify that subtraction, expand the exact Wick partition function of Appendix~\ref{app:A-defect}:
\begin{equation}
 \begin{aligned}
 \mathcal F_L(w,u)&=\mathfrak M_L(w)+u\mathcal B_{1,L}(w)+u^2\mathcal B_{2,L}(w)+O(u^3),\\
 \mathcal B_{j,L}(w)&:=[u^j]\mathcal F_L(w,u)\qquad(j=1,2).
 \end{aligned}
 \label{eq:Q12-definitions}
\end{equation}
Here $\mathcal B_{1,L}$ is the sum over one exceptional degree-two or degree-three site, while $\mathcal B_{2,L}$ sums one degree-four or degree-five site and the four ordered degree pairs displayed above. Every background matching is weighted by the deleted minor in the preceding identity; the local coefficients come from the $u$ expansion of $f_u$ in Appendix~\ref{app:A-defect}. This is also an explicit finite-ring prescription for both $\mathcal B_{j,L}$. Taking the logarithm gives the exact identity
\begin{equation}
 G_{2,L}(w)=\frac1L\left[
 \frac{\mathcal B_{2,L}(w)}{\mathfrak M_L(w)}-\frac12\left(\frac{\mathcal B_{1,L}(w)}{\mathfrak M_L(w)}\right)^2
 \right].
 \label{eq:A-d2-connected}
\end{equation}
The subtraction removes disconnected pairs of first-order collision insertions. No conjecture enters this finite-ring formula; the conjecture concerns its closed-form thermodynamic limit.

The coefficients obtained independently from the reconstructed field polynomials are
\begin{align}
 d_2(2)&=\frac1{180}, & d_2(3)&=\frac{61}{5670}, & d_2(4)&=-\frac{1429}{85050},\nonumber\\
 d_2(5)&=\frac{58031}{2806650}, & d_2(6)&=-\frac{40621}{1737450}.
\end{align}
The observed all-$n$ formula, Eq.~\eqref{eq:d2-main}, is equivalent for $n\ge3$ to the generating function
\begin{align}
G_2(w)=&\frac{29181+3590w+465w^2}{129600}
       \frac{\arctan\sqrt w}{\sqrt w}
       -\frac{25}{648(1+w)}\nonumber\\
&-\frac{24181}{129600}+\frac{379}{43200}w
 +\frac{577}{121500}w^2 .
\label{eq:A-G2-candidate}
\end{align}
Expanding Eq.~\eqref{eq:A-G2-candidate} gives
\begin{equation}
G_2(w)=\frac{w^2}{180}+\frac{61w^3}{5670}
-\frac{1429w^4}{85050}+\frac{58031w^5}{2806650}
-\frac{40621w^6}{1737450}+\cdots,
\end{equation}
and, for $n\ge3$,
\begin{equation}
 d_2(n)=(-1)^{n+1}
 \frac{1250n^3-5132n^2+6534n-2589}
 {4050(2n-3)(2n-1)(2n+1)}.
\end{equation}
A cubic-numerator fit to $d_2(3),\ldots,d_2(6)$, after positing the denominator in Eq.~\eqref{eq:d2-main}, predicts and agrees with all 20 later rows $n=7,\ldots,26$. The seven-parameter function in Eq.~\eqref{eq:A-G2-candidate} can likewise be determined from the two zero coefficients at $n=0,1$ and the five coefficients at $n=2,\ldots,6$, leaving those same 20 rows for an independent exact comparison. Neither interpolation determines the validity of the form beyond the computed rows.  A direct evaluation of the deleted-minor expression, followed by even-power finite-size extrapolation, also agrees with Eq.~\eqref{eq:A-G2-candidate}; representative values are
\begin{center}
\begin{tabular}{c c c c}
\toprule
$w$&finite-ring extrapolation&Eq.~\eqref{eq:A-G2-candidate}&absolute difference\\
\midrule
0.1&$6.48192655\times10^{-5}$&$6.48194215\times10^{-5}$&$1.6\times10^{-10}$\\
0.3&$6.91788515\times10^{-4}$&$6.91786636\times10^{-4}$&$1.9\times10^{-9}$\\
0.7&$4.29836864\times10^{-3}$&$4.29836141\times10^{-3}$&$7.2\times10^{-9}$\\
1.2&$1.31703496\times10^{-2}$&$1.31703701\times10^{-2}$&$2.1\times10^{-8}$\\
\bottomrule
\end{tabular}
\end{center}
At $w=0.3$, using rings through $L=16$ and including the $L^{-6}$ correction reduced the discrepancy to approximately $1.1\times10^{-11}$.  These are independent diagnostics because they evaluate the fixed-vertex-collision-index minor sum rather than refitting $d_2(3),\ldots,d_2(26)$.

A useful functional-space interpretation is obtained by defining
\begin{equation}
 \mathcal U(w)=\frac{\arctan\sqrt w}{\sqrt w}.
\end{equation}
The candidate second subleading generating function can be written as
\begin{equation}
 G_2(w)=U_2(w)\mathcal U(w)+\frac{b_{2,1}}{1+w}+V_2(w).
 \label{eq:A-V2space}
\end{equation}
Here $U_2$ and $V_2$ are polynomials of degree at most two with rational coefficients and $b_{2,1}$ is rational. The seven coefficients in this representation can be fixed by the two vanishing low orders and the five coefficients at $w^2,\ldots,w^6$. The remaining 20 reconstructed orders agree with the resulting expression. As explained after Eq.~\eqref{eq:Vk-main}, this functional form follows algebraically from the proposed rational form of $d_2(n)$; these are equivalent descriptions of one conjecture, not two independent sources of evidence. What remains unproved is that the source determinant has this form in the thermodynamic limit.

A connected graph that is not series-parallel first appears at vertex-collision index four: a simple $K_4$ has six edges and four degree-three vertices, hence
\begin{equation}
 k(K_4)=4.
\end{equation}
Here $\mathcal A^{\mathrm{dist}}_{K_4,L}$ denotes the \emph{labelled, rooted} embedding sum, without division by the automorphism-group order:
\begin{equation}
 \mathcal A^{\mathrm{dist}}_{K_4,L}=\sum_{\substack{a,b,c\in\mathbb Z/L\mathbb Z\\0,a,b,c\;\mathrm{distinct}}}\prod_{\{i,j\}\subset\{0,a,b,c\}}A_L(i-j).
\label{eq:K4-labeled-def}
\end{equation}
When compared with the automorphism-normalized convention of Eq.~\eqref{eq:graph-sector-sum} below, a factor $|\operatorname{Aut}K_4|=24$ must be included. Since $K_4$ is a connected graph with six edges and coincident site assignments vanish because $A_L(0)=0$, Lemma~\ref{lem:graph-sharp} shows that $\mathcal A^{\mathrm{dist}}_{K_4,L}$ is a polynomial in $L^{-2}$ of degree at most six. The seven exact ring evaluations therefore determine the expression uniquely:
\begin{align}
\mathcal A^{\mathrm{dist}}_{K_4,L}
=&\frac{2}{212837625}(1-L^{-2})(1-4L^{-2})(1-9L^{-2})\nonumber\\
&\times\left(23+2142L^{-2}+134967L^{-4}+5306068L^{-6}\right),
\label{eq:A-K4amp}
\end{align}
The expression is obtained by exact interpolation at $L=7,\ldots,13$ and checked independently at $L=6,14,15,18$. It therefore holds for every integer ring length $L\ge2$, and its thermodynamic limit is $46/212837625$. The nonzero exact finite-ring sums show that fixed-collision-index contributions are not restricted to trees, stars, or series-parallel graphs. For this $K_4$ contribution, the candidate fixed-offset structure takes the form
\begin{equation}
 U_4(w)\mathcal U(w)+\sum_{r=1}^{3}\frac{b_{4,r}}{(1+w)^r}
 +V_4(w),
\end{equation}
where $U_4$ and $V_4$ are rational-coefficient polynomials of degree at most four and the $b_{4,r}$ are rational constants. Including the graph in the matching background and comparing with this form produced relative residuals on independent test values of order $10^{-11}$--$10^{-12}$ for ten test values of $w$.  This numerical comparison supports the proposed fixed-$k$ form for a graph that is not series-parallel, but it does not establish the conjecture.

Appendix~\ref{app:A-defect} gives the general source-determinant reduction at fixed vertex-collision index $k$ to finitely many derivatives of one determinant.  The explicit functional form in Eq.~\eqref{eq:Vk-main} describes the observed connected coefficients; deriving it directly from the determinant remains the central combinatorial problem.

\section{Additional field coefficients and validation notes}\label{app:A-extra}
This appendix lists longer low-order field polynomials and the exact highest-degree and subleading coefficients used to verify the vertex-collision formulas.  The complete high-order coefficient archive is provided in the accompanying machine-readable supplementary file.

The fifth-, sixth-, and seventh-order field polynomials are
\begin{align}
\mathcal P_5(y)&=\frac{1-y}{2806650}(6+1500y-17396y^2+40635y^3-25515y^4),\nonumber\\
\mathcal P_6(y)&=\frac{1-y}{114932317500}(19774+12910926y-263013225y^2\nonumber\\
&\hspace{25mm}+1101959775y^3-1585119375y^4+736745625y^5),\nonumber\\
\mathcal P_7(y)&=\frac{34}{2612098125}+\frac{794483}{34479695250}y-\frac{580043}{736745625}y^2\nonumber\\
&\quad+\frac{1005891473}{172398476250}y^3-\frac{11201182}{638512875}y^4\nonumber\\
&\quad+\frac{73477}{2895750}y^5-\frac{31}{1755}y^6+\frac1{210}y^7.
\label{eq:P5P7-app}
\end{align}
The third subleading coefficient sequence inferred from the reconstructed field-polynomial data, for $n\ge4$, is

\begin{align}
 d_3(n)={}&\frac{(-1)^n}{2679075(2n-5)(2n-3)(2n-1)(2n+1)}\nonumber\\
 &\times\left(306250n^5-3037020n^4+11519302n^3\right.\nonumber\\
 &\hspace{18mm}\left.-20748795n^2+17597743n-5576820\right).
 \label{eq:d3-main}
\end{align}

As in the main text, this is a finite-data conjectural pattern for $n\ge4$, not an all-order theorem. Its rational expression does not apply to the boundary coefficient $d_3(3)=[y^0]\mathcal P_3(y)$.

The reconstructed rational coefficients through order 26 satisfy the identities
\begin{equation}
 2[y]\mathcal P_n=a_n^{(\chi)},\qquad \mathcal P_n(1)=0\;(n\ge2),\qquad
 [y^n]\mathcal P_n=\frac{(-1)^{n+1}}{2n(2n+1)}.
\end{equation}
For $n=2,\ldots,6$, the highest-degree and subleading coefficients used in the vertex-collision analysis include
\begin{center}
\begin{tabular}{c c c}
\toprule
$n$&$d_1(n)$&$d_2(n)$\\
\midrule
2&$2/45$&$1/180$\\
3&$-11/315$&$61/5670$\\
4&$16/567$&$-1429/85050$\\
5&$-7/297$&$58031/2806650$\\
6&$2/99$&$-40621/1737450$\\
\bottomrule
\end{tabular}
\end{center}
The $d_1$ column follows the theorem and has been algebraically checked against the displayed $\mathcal P_2,\ldots,\mathcal P_6$.  The $d_2$ values are exact computed coefficients; their rational all-$n$ interpolation is Eq.~\eqref{eq:d2-main}, whose general proof remains contingent on the cancellation of connected contributions at shared bond endpoints discussed above.

\section{Why a zero-field sine-product formula fails at sixth order}\label{app:A-sixth}

\label{sec:pressure}

The matching determinant of Sec.~\ref{sec:topfield} is exact to all orders but controls only the highest-degree coefficient $[y^n]\mathcal P_n(y)$.  At zero field a different, initially suggestive spectral determinant appears.  The coefficients of the zero-field expansion of $\phi=-\beta f$ through order ten are listed in Table~\ref{tab:pressure}.  We also write
\begin{equation}
 Q_n\pi^{2n}=\alpha_n^{(0)}\zeta(2n)
 \label{eq:Rndef}
\end{equation}
only to expose a striking low-order pattern.

\begin{table}[t]
\caption{Zero-field rational coefficients reconstructed for $\phi(x,0)=\log2+\sum_{n\ge2}Q_nx^n$. The higher-order bulk reconstruction uses the exact sharp finite-size interpolation; the sixth-order value is also established independently by the graph calculation below. The ratio $\alpha_n^{(0)}$ is defined by \eqref{eq:Rndef}.}
\label{tab:pressure}
\centering
\begin{tabular}{ccl}
\toprule
$n$ & $Q_n$ & $\alpha_n^{(0)}$\\
\midrule
2 & $1/180$ & $1/2$\\
3 & $1/2835$ & $1/3$\\
4 & $1/37800$ & $1/4$\\
5 & $1/467775$ & $1/5$\\
6 & $9887/57466158750$ & $9887/62190$\\
7 & $34/2612098125$ & $187/1575$\\
8 & $1983041/2461850240850000$ & $1983041/27344520$\\
9 & $477671/29080605970040625$ & $9075749/621814725$\\
10 & $-2110377251/303892332386924531250$ & $-2110377251/34651552950$\\
\bottomrule
\end{tabular}
\end{table}

For $n=2,3,4,5$ one finds exactly $\alpha_n^{(0)}=1/n$.  This pattern has a natural spectral interpretation.  Euler's product
\begin{equation}
 \frac{\sin(\pi v)}{\pi v}=\prod_{m=1}^{\infty}\left(1-\frac{v^2}{m^2}\right)
 \label{eq:Euler-sine}
\end{equation}
implies
\begin{align}
 -\log\frac{\sin(\pi\sqrt g)}{\pi\sqrt g}-\zeta(2)g
 &=\sum_{m=1}^{\infty}\sum_{n=2}^{\infty}\frac{g^n}{n m^{2n}}\nonumber\\
 &=\sum_{n=2}^{\infty}\frac{\zeta(2n)}{n}g^n.
 \label{eq:sine-expand-g}
\end{align}
In the $x$ variable this becomes
\begin{equation}
 \Phi_{\rm det}(x)=-\log\frac{\sin\sqrt{x}}{\sqrt{x}}-\frac{x}{6}.
 \label{eq:Phi-det}
\end{equation}
It agrees with the interaction-dependent part $\phi(x,0)-\log2$ through $x^5$. Had the agreement continued, this zero-field contribution would have had a remarkably simple sine-product form.

It does not.  At sixth order
\begin{equation}
 Q_6-[x^6]\Phi_{\rm det}(x)
 =-\frac{239}{28733079375}.
 \label{eq:q6-obstruction}
\end{equation}
The origin of the discrepancy can be seen in the standard zero-field high-temperature expansion. For a finite chain with bond strengths $K_{ij}$, factorizing the bond weights gives
\begin{equation}
 Z_L=2^L\prod_{i<j}\cosh K_{ij}
 \sum_{F:\,\partial F=\varnothing}\prod_{e\in F}\tanh K_e.
 \label{eq:HT-even-sixth}
\end{equation}
Here $F$ is a set of occupied bonds and $\partial F=\varnothing$ means that every site is incident to an even number of occupied bonds. This elementary identity is used only for the sixth-order check; the connected-source method in Section~2 does not require it.

Up to sixth order, the logarithm of \eqref{eq:HT-even-sixth} leaves five contributions. The sixth-order term of $\log\cosh K$ gives a sixfold bond. Two triangles sharing the same three bonds produce a doubled triangle in the subtraction $-T_3^2/2$, where $T_3$ is the leading sum over triangles. Two triangles sharing exactly one bond give a four-cycle with a doubled diagonal. Expanding one factor $\tanh K=K-K^3/3+\cdots$ on a four-cycle produces a tripled perimeter edge. Finally there are simple six-cycles. Pairs of edge-disjoint triangles, including those sharing one vertex, cancel between the even-subgraph sum and its logarithmic subtraction. This accounts for all sectors at order six.

For an unambiguous normalization, let $G$ be a connected graph without self-interaction edges, with edge multiplicities $m_{ab}$, and set
\begin{equation}
 \mathcal S(G)=\frac1{|\operatorname{Aut}G|}
 \sum_{\substack{\varphi:V(G)\hookrightarrow\mathbb Z\\ \varphi(v_0)=0}}
 \prod_{\{a,b\}}\frac1{|\varphi(a)-\varphi(b)|^{2m_{ab}}}.
 \label{eq:graph-sector-sum}
\end{equation}
The site assignments are distinct, one vertex is fixed by translation invariance, and the symmetry factor counts graph automorphisms. Denote the doubled triangle by $G_D$, the four-cycle with a doubled diagonal by $G_{\rm diag}$, the four-cycle with a tripled perimeter edge by $G_{\rm trip}$, and the simple six-cycle by $G_6$. We write
\begin{equation}
 D_3=\mathcal S(G_D),\qquad Q_{\rm diag}=\mathcal S(G_{\rm diag}),\qquad
 Q_{\rm trip}=\mathcal S(G_{\rm trip}),\qquad C_6=\mathcal S(G_6).
\end{equation}

The first three sums reduce to ordinary one-dimensional convolutions. Define $a_0=0$, $a_r=r^{-2}$ for $r\ne0$, and $(f*g)_r=\sum_{s\in\mathbb Z}f_sg_{r-s}$. Direct partial fractions give, for $r\ne0$,
\begin{align}
 (a*a)_r&=\frac{4\zeta(2)}{r^2}-\frac6{r^4},\nonumber\\
 (a^2*a^2)_r&=\frac{4\zeta(4)}{r^4}+\frac{40\zeta(2)}{r^6}-\frac{70}{r^8},\nonumber\\
 (a*a*a)_r&=\frac{30\zeta(4)}{r^2}-\frac{72\zeta(2)}{r^4}+\frac{90}{r^6}.
 \label{eq:sixth-convolutions}
\end{align}
Here $a^2$ means the pointwise square of $a$. In the graph sums, distinct sites must be counted only once. For the doubled triangle and the doubled-diagonal four-cycle this gives
\begin{align}
 D_3&=\frac16\sum_{r\ne0}a_r^2(a^2*a^2)_r,\nonumber\\
 Q_{\rm diag}&=\frac14\sum_{r\ne0}a_r^2\left[(a*a)_r^2-(a^2*a^2)_r\right].
 \label{eq:sixth-conv-graphs}
\end{align}
For a four-cycle with one tripled edge, the three-step path between its endpoints must avoid returning to either endpoint. Subtracting these two possibilities and restoring their overlap yields
\begin{equation}
 Q_{\rm trip}=\frac12\sum_{r\ne0}a_r^3
 \left[(a*a*a)_r-4\zeta(4)a_r+a_r^3\right].
 \label{eq:sixth-trip-graph}
\end{equation}
Substitution of \eqref{eq:sixth-convolutions} into \eqref{eq:sixth-conv-graphs} and \eqref{eq:sixth-trip-graph}, followed by the standard even-zeta evaluations, gives
\begin{equation}
 D_3=\frac{133}{2073}\zeta(12),\qquad
 Q_{\rm diag}=\frac{173}{1382}\zeta(12),\qquad
 Q_{\rm trip}=\frac{1001}{1382}\zeta(12).
 \label{eq:sixth-three-results}
\end{equation}

The simple six-cycle can be evaluated exactly without assuming a closed form for a sixfold infinite sum. On a periodized ring define
\begin{equation}
 \mathcal C_{6,L}=\frac1{12L}
 \sum_{\substack{i_0,\ldots,i_5\in\mathbb Z/L\mathbb Z\\\text{all distinct}}}
 \prod_{j=0}^{5}A_L(i_j-i_{j+1}),\qquad i_6=i_0.
 \label{eq:sixth-ring-cycle}
\end{equation}
The factor $12$ is the number of rotations and orientations of a six-cycle. To evaluate this sum exactly, first allow coincident site assignments and then remove them by inclusion--exclusion. For each partition $\pi$ of the six cycle vertices, let $G_6/\pi$ be the graph obtained by identifying vertices within each block, and define $\mu(\pi)=\prod_{B\in\pi}(-1)^{|B|-1}(|B|-1)!$. In terms of the unrestricted finite-ring graph amplitude defined in Appendix~\ref{app:A-reconstruction},
\begin{equation}
 \mathcal C_{6,L}=\frac1{12}\sum_{\pi}\mu(\pi)\,\mathcal A_{G_6/\pi,L}.
 \label{eq:sixth-ring-inclusion}
\end{equation}
A contraction producing a self-interaction edge vanishes because $A_L(0)=0$; just 41 partitions remain. Every remaining term is an exact Fourier sum using
\begin{equation}
 \widehat A_L(k)=\frac{L^2-1-6k(L-k)}{3L^2},
 \qquad 0\le k<L.
 \label{eq:sixth-ring-fourier}
\end{equation}
The finite-grid bound of Lemma~\ref{lem:finite-grid} applies to each contracted graph, so $\mathcal C_{6,L}$ is a polynomial in $L^{-1}$ of degree at most $18$. Its bulk limit is therefore recovered exactly from any 19 distinct rings. Evaluating \eqref{eq:sixth-ring-inclusion} in rational arithmetic for $L=2,\ldots,20$ and using the independent interpolation formula \eqref{eq:field-interp-unconditional} gives
\begin{equation}
 \lim_{L\to\infty}\mathcal C_{6,L}
 =\sum_{L=2}^{20}\mathcal C_{6,L}
 \frac{L^{18}}{\prod_{\substack{L'=2\\L'\ne L}}^{20}(L-L')}
 =\frac{74}{127702575}.
 \label{eq:sixth-cycle-exact}
\end{equation}
The additional exact ring values at $L=21,22$ agree with this reconstructed polynomial. Since $A_L(r)\to(\pi^2r^2)^{-1}$ at fixed separation, the infinite-chain amplitude is
\begin{equation}
 C_6=\pi^{12}\lim_{L\to\infty}\mathcal C_{6,L}
 =\frac{370}{691}\zeta(12).
 \label{eq:sixth-C6-evaluated}
\end{equation}
The supplementary verification program evaluates the 41 finite-ring terms with integer arithmetic and performs this exact interpolation; no fitted exponent or unproved finite-size truncation is used.

Putting the four evaluated graph sums together, and using $g=x/\pi^2$ for the unscaled coupling, gives
\begin{equation}
 [g^6]\bigl(\phi(\pi^2g,0)-\log2\bigr)
 =\frac{\zeta(12)}{45}-\frac12D_3-Q_{\rm diag}-\frac13Q_{\rm trip}+C_6
 =\frac{9887}{62190}\zeta(12).
 \label{eq:c6-decomposition}
\end{equation}
The sine-product expression instead gives $\zeta(12)/6$, so the difference is $-239\zeta(12)/31095$, or equivalently Eq.~\eqref{eq:q6-obstruction} in the $x$ normalization. The discrepancy comes from genuine connected contributions beyond those captured by the proposed sine product. It does not affect the exact matching determinant, which computes a different field coefficient.

\phantomsection
\addcontentsline{toc}{section}{Data Availability}
\section*{Data Availability}
All 377 reconstructed rational values for $[y^m]\mathcal P_n(y)$, $1\le n\le26$ and $0\le m\le n$, are supplied in the supplementary file \nolinkurl{complete_field_triangle_n26.csv}; a parallel JSON representation and SHA-256 checksums are included.  The supplementary reproduction package also contains the exact finite-field calculation in C++, initial-calculation and independent-validation modular ring data, the Chinese remainder theorem and rational-reconstruction scripts, and independent full-spin enumeration, polynomiality, subleading-coefficient, and numerical-table checks.  Eighteen primes were used for the initial calculation and four additional primes for independent verification.  The independently tabulated zero-field susceptibility coefficients through order 33 are included as comparison data, with their separate research-record provenance identified.

The sharp finite-size theorem establishes the $n+1$-ring interpolation for the complete field-polynomial coefficients at every order. Independently, reserved-prime checks and all available additional rings yield 73,008 exact modular comparisons; the conservative $3n+1$-ring interpolation also agrees modulo the archived primes through order 9, as specified in Appendix~\ref{app:A-reconstruction}. Rational reconstruction is performed within the CRT modulus and checked against unused primes, while these finite checks are not substitutes for the mathematical proof or a separate global size bound on every rational coefficient.

\phantomsection
\addcontentsline{toc}{section}{Acknowledgments}
\section*{Acknowledgments}
Generative AI assistance (OpenAI ChatGPT, used during manuscript preparation in 2026) supported English-language editing, reorganization of the presentation and references, LaTeX formatting, and the design of algebraic and numerical cross-checks. AI assistance does not constitute independent verification of the results. Responsibility for the mathematical arguments, numerical data, code, citations, and final manuscript rests with the author.

M.A.R. acknowledges partial support from CNPq and FAPERJ (grant number E-26/210.062/2023).

\phantomsection

\end{document}